\documentclass{LMCS}
\usepackage{array,amssymb,graphicx,enumerate,hyperref}

\newcommand{\cL}{\mathcal{L}}
\newcommand{\Fml}{\mathsf{Fml}}
\newcommand{\bin}{\mathsf{bin}}

\newenvironment{bDlist}{\begin{trivlist}}{\end{trivlist}}
\newcommand{\bDitem}{\item $\blacktriangleright$\hspace*{1mm}}

\def\doi{5 (3:3) 2009}
\lmcsheading%
{\doi}
{1--27}
{}
{}
{Feb.~16, 2008}
{Aug.~\phantom{0}3, 2009}
{} 

\begin{document}
\title[Lindstr{\"o}m theorems for fragments of first-order
  logic]{Lindstr{\"o}m theorems for fragments of first-order
  logic\rsuper *}
\author[J.~van Benthem]{Johan van Benthem\rsuper a}
\address{{\lsuper{a,c}}ILLC, University of Amsterdam}
\email{\{johan,vaananen\}@science.uva.nl}

\author[B.~ten Cate]{Balder ten Cate\rsuper b}
\address{{\lsuper b}ISLA, University of Amsterdam}
\email{balder.tencate@uva.nl}

\author[J.~V\"a\"an\"anen]{Jouko V\"a\"an\"anen\rsuper c}
%\address{{\lsuper c} ILLC, University of Amsterdam}
%\email{vaananen@science.uva.nl}

\titlecomment{{\lsuper *}An extended abstract of this paper
  \cite{UsLICS} has appeared in the proceedings of LICS 2007. We would
  like to thank Martin Otto and the anonymous referees for helpful
  comments. This work is partially supported by the Netherlands
  Organization for Scientific Research (NWO) grant 639.021.508.}

\begin{abstract}
   Lindstr\"om theorems characterize logics in terms of
   model-theoretic conditions such as Compactness and the
   L\"owenheim-Skolem property. Most existing characterizations of
   this kind concern extensions of first-order logic. But on the other
   hand, many logics relevant to computer science are fragments or
   extensions of fragments of first-order logic, e.g., $k$-variable
   logics and various modal logics. Finding Lindstr\"om theorems for
   these languages can be % challenging, as most known techniques rely
   on coding arguments that seem to require the full expressive power
   of first-order logic.

   In this paper, we provide Lindstr\"om theorems for several
   fragments of first-order logic, including the $k$-variable
   fragments for $k>2$, Tarski's relation algebra, graded modal logic,
   and the binary guarded fragment. We use two different proof
   techniques. One is a modification of the original Lindstr\"om
   proof.  The other involves the modal concepts of bisimulation, tree
   unraveling, and finite depth. Our results also imply semantic
   preservation theorems.
\end{abstract}

\maketitle\vfill

\section{Introduction}

There are many ways to capture the expressive power of a logical
language $\cL$. For instance, one can characterize $\cL$ as being a
model-theoretically well behaved fragment of a richer language $\cL'$
(a \emph{preservation theorem}), or as being maximally expressive
while satisfying certain model-theo\-retic properties (a
\emph{Lindstr\"om theorem}).  The main contribution of this paper is a
series of Lindstr\"om theorems for fragments of first-order logic. We
also show connections between our Lindstr\"om theorems and
preservation theorems.

The original Lindstr\"om theorem for first-order logic, in one of its 
most widely used
formulations, says the following:
\eject\vfill\relax

\begin{trivlist}\item \textbf{The first-order Lindstr\"om Theorem 
\cite{Lindstroem69:extensions}}
   {\em An extension of first-order logic satisfies Compactness and the
   L\"owenheim-Skolem property if and only if it is no more expressive than
   first-order logic.}
\end{trivlist}

\begin{figure}
\begin{center}
\includegraphics[scale=0.85]{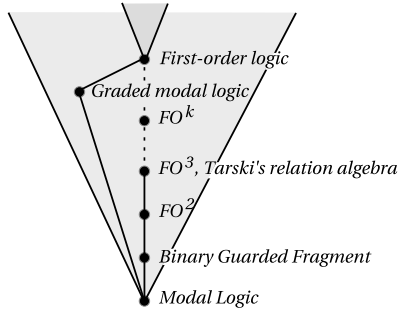}
\end{center}
\caption{Fragments of first-order logic}
\label{fig:hierarchy}
\end{figure}
There are several other versions of the theorem, characterizing
first-order logic for instance in terms of Compactness and invariance
for potential isomorphisms (the 'Karp Property'). Analogues of this 
result have been
obtained for various extensions of first-order logic.
On the other hand, few Lindstr\"om theorems are known for
\emph{fragments} of first-order logic. One notable example
is Van Benthem's recent Lindstr\"om theorem for modal logic:
\begin{trivlist}\item \textbf{The modal Lindstr\"om theorem
     \cite{benthem07:lindstrom}} \em An extension of basic modal logic
   satisfies Compactness and Bisimulation Invariance if and only if it is no more
   expressive than basic modal logic. \footnote{By \emph{extensions of
     basic modal logic} we mean language extensions, not
     axiomatic extensions.}
\end{trivlist}

We are not aware of a similar characterization for modal logic involving the
L\"owenheim-Skolem property.  Note that first-order logic itself is a
compact proper extension of modal logic that has the latter property.

Our primary motivation for considering fragments comes from computer
science logic.  Many logics relevant to computer science are fragments
(or non-FO extensions of fragments) of first-order logic, for example
$k$-variable logics and various modal logics. Finding Lindstr\"om
theorems for such languages can be a challenging problem, since most
techniques used in the past to prove Lindstr\"om theorems rely on
coding arguments that seem to require the full expressive power of
first-order logic. For a recent survey of Lindstr\"om theorems in a
general setting, see \cite{garciamatos05:abstract}.  Another
motivation is that, by widening our domain of study to logics not
necessarily extending first-order logic, we may come to know more
about first-order logic itself.

We follow two global lines of attack. First, we take the original
Lindstr\"om theorem for first-order logic and generalize the
proof as much as possible. In this way, we obtain Lindstr\"om theorems
for the finite variable fragments $FO^k$ with $k>2$ and Tarski's
relation algebra. Next, we take the modal Lindstr\"om theorem as a
starting point, and try to generalize it to richer languages. In this
way, we obtain Lindstr\"om theorems for graded modal logic (on
arbitrary Kripke structures and on trees) and the binary guarded
fragment.

Many open questions remain. For example, we have not been able to
find Lindstr\"om theorems for the two-variable fragment or the
full guarded fragment.

\section{From first-order logic downwards}

In this first part, we take the classic Lindstr\"om theorem as a
starting point, and we show that the proof generalizes to certain
fragments of first-order logic. 

\subsection{A strengthening of the Lindstr\"om theorem for first-order
   logic over binary vocabularies}

The first-order Lindstr\"om theorem uniquely characterizes first-order
logic in terms of Compactness and the L\"owenheim-Skolem property
\emph{within the class of all its extensions}. As we will show in this
section, this result can be improved: first-order logic can be
characterized in terms of Compactness and the L\"owenheim-Skolem
property \emph{within the class of all extensions of the
   three-variable fragment $FO^3$}, if we consider vocabularies
consisting only of unary and binary relation symbols. The proof is not
substantially more difficult than that of the original Lindstr\"om
theorem, requiring mainly some extra care with coding partial 
isomorphisms and back-and-forth behaviour. But this simple strengthening 
will allow us to obtain a number of new
results on Tarski's relation algebra, as well as
finite variable fragments.

To keep things simple, we will work with a fixed relational signature
consisting of a set of unary relation symbols and a set of binary
relation symbols, both countably infinite.

By an \emph{abstract logic} we mean a pair
$\cL=(\Fml_{\cL},\models_\cL)$, where $\Fml_{\cL}$ is the set of
sentences of $\cL$ and $\models_{\cL}$ is a binary truth relation between
$\cL$-sentences and models. If no confusion arises, we will sometimes 
write $\cL$
for $\Fml_{\cL}$ and $\models$ for $\models_{\cL}$.  We assume that
$\cL$-sentences are preserved under isomorphisms, and that $\cL$ has
the following properties:
\begin{bDlist}
\bDitem \emph{closure under Boolean connectives}:
  for every $\phi\in\cL$ there is a sentence $\psi\in\cL$ expressing its
negation (i.e., for all models $M$, $M\models\psi$ iff
$M\not\models\phi$), and for every $\phi,\psi\in\cL$ there
is a sentence $\chi\in\cL$ expressing the conjunction of $\phi$ and $\psi$.
\bDitem \emph{closure under renamings}: for every mapping $\rho$
sending relation symbols to relation symbols of the same arity, and
for every sentence $\phi\in\cL$, there is a sentence $\psi\in\cL$ such
that for all models $M$, $M\models\psi$ iff
$\rho(M)\models\phi$.
\bDitem \emph{closure under relativisation
   by unary predicates}: for every sentence $\phi\in\cL$
and unary relation symbol $P$, there is a sentence $\psi\in\cL$ such
that for all models $M$, $M\models\psi$ iff $M^P\models\phi$, with
$M^P$ the submodel of $M$ induced by the subset defined by $P$. 
\end{bDlist}
Examples of abstract logics in this sense include first-order logic 
($FO$) and its
$k$-variable fragments ($FO^k$), with $k\geq 1$.

Given two abstract logics, $\cL$ and $\cL'$, we say that $\cL$ extends
$\cL'$ (or, $\cL'$ is contained in $\cL$, denoted by $\cL'\subseteq
\cL$), if there is a map $f: \Fml_{\cL'} \to \Fml_{\cL}$ preserving
truth in the sense that, for all models $M$ and sentences $\phi\in\cL'$,
$M\models_{\cL'}\phi$ iff $M\models_{\cL} f(\phi)$.

An abstract logic $\cL$ has Compactness if for every set of
$\cL$-formulas $\Sigma$, if every finite subset of $\Sigma$ is
satisfiable then the entire set $\Sigma$ is satisfiable. An abstract
logic $\cL$ has the L\"owenheim-Skolem property if every satisfiable
set of $\cL$-formulas has a countable model.

First, we show that each compact extension of $FO$
(in fact, already of $FO^2$) has the
``finite occurrence property''.

\begin{lem}[Finite occurrence property]\label{lem:finite-occurrence-FO}
   Let $\cL$ be any abstract logic extending $FO^2$ that has
   Compactness.  Then for any $\phi\in\cL$ there is a finite
   set of relation symbols $REL(\phi)$ such that the truth of $\phi$ in
   any model is independent of the denotation of relation symbols
   outside $REL(\phi)$.
\end{lem}

\begin{proof}
   A standard argument, cf.~\cite{Flum85:characterizing}:
   since our vocabulary contains infinitely many unary and binary
   relation symbols, there are renamings $\rho_1, \rho_2$ whose range
   is disjoint. Take any $\phi\in\cL$, and let $\phi_1, \phi_2$ be its
   renamings  according to $\rho_1$ and $\rho_2$. Let $\Sigma
   = \{\forall x_1\ldots x_k(\rho_1(R)(x_1\ldots x_k)\leftrightarrow 
\rho_2(R)(x_1\ldots x_k))\mid
   \text{$R$ a $k$-ary relation symbol}\}$. Then $\Sigma\models
   \phi_1\leftrightarrow\phi_2$, and hence, by Compactness, a finite
   subset $\Sigma'\subseteq\Sigma$ implies
   $\phi_1\leftrightarrow\phi_2$. We can pick for
   $REL(\phi)$ the relation symbols occurring in $\Sigma'$.
   Note that, in case of binary vocabularies, all formulas in $\Sigma$
   belong to $FO^2$.
\end{proof}

The key to the proof of our improved Lindstr\"om theorem is the
following observation:

\begin{lem}
   \label{lem:proj-def-finiteness}
   Let $\cL$ be any abstract logic with the L\"owenheim-Skolem
   property and the Finite Occurrence Property, such that $\cL$ extends $FO^3$ and is not contained in
   $FO$.  Then ``$\cL$ can relatively projectively define finiteness'':
   there is a formula $\psi\in \cL$ with a unary predicate $N$
   such that, for each $n\in\mathbb{N}$, there is a model of $\psi$
   with exactly $n$ elements satisfying $N$, while no model of $\psi$ has
   infinitely many elements satisfying $N$. 
\end{lem}

\begin{proof}
   The basic idea is the same as in traditional proofs of the
   Lindstr\"om theorem (e.g., \cite{Flum85:characterizing}). Our main
   contribution is to show that, in the case of binary vocabularies,
   the coding argument requires only three variables.

   Take any $\phi\in \cL$ not belonging to $FO$. Then for each $k\in
   \mathbb{N}$, there are models $\mathfrak{A}_k\models\phi$ and
   $\mathfrak{B}_k\not\models\phi$ that are potentially isomorphic up
   to back-and-forth depth $k$. At the same time, no potentially
   isomorphic models can disagree on the sentence $\phi$. We can describe 
this situation
   \emph{inside} $\cL$. The construction is outlined in
   Figure~\ref{fig:lindstrom3var}. 

\begin{figure}[t]
\begin{center}
\includegraphics[width=0.5\linewidth]{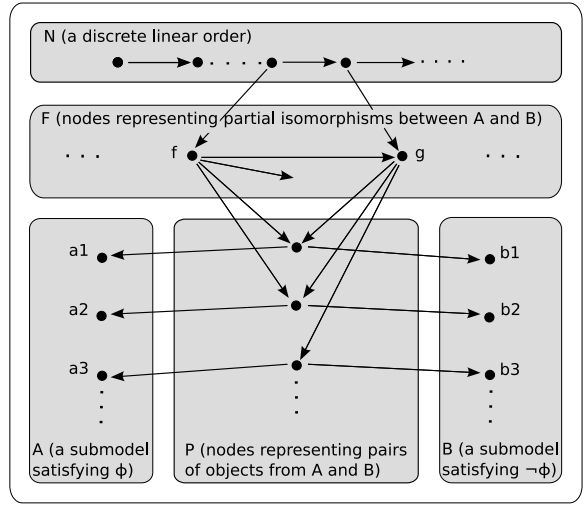}
\end{center}
\caption{Model from the proof of Lemma~\ref{lem:proj-def-finiteness}.}
\label{fig:lindstrom3var}
\end{figure}

The model depicted in Figure~\ref{fig:lindstrom3var} describes
two models, connected via a collection of
partial isomorphisms, that disagree on the sentence $\phi$. The most important
feature is that, if $N$ is an infinite set, then the collection of
partial isomorphisms constitutes a potential isomorphism, whereas if
$N$ is finite (say, of size $k$),  the
collection of partial isomorphisms constitutes a potential isomorphism
\emph{up to back-and-forth depth $k$}.

More precisely, $A$ and $B$ are unary predicates defining the domains
of two (sub)models, $P$ is a unary predicate whose elements denote
pairs from $A\times B$, and the elements of $F$ represent partial
isomorphisms (i.e., sets of pairs constituting structure preserving
bijections). The arrows represent a binary relation $R$. For instance,
in the given example, $f$ represents the partial isomorphism
$\{(a_1,b_1),(a_2,b_2)\}$, and $g$ represents the partial
isomorphism that extends $f$ with the pair $(a_3, b_3)$. The elements
of the linearly ordered set $N$ will be used to associate a finite 
index to each partial isomorphism. 

\begin{trivlist}\item \textbf{Claim:}
Each of the following properties of this model can be expressed by a
sentence of $\mathcal{L}$:

\begin{enumerate}[(1)]%\setlength{\itemsep}{0mm}
\item Every $p\in P$ is associated to a pair from $A\times B$.

\item Every $f\in F$ is associated to a set of elements of $P$ that
   form a partial \emph{bijection} between $A$ and $B$.

\item Each such partial bijection preserves structure on the submodels
   defined by $A$ and $B$, as far as the finitely many relations
   occurring in $\phi$ are concerned (recall that $\cL$ has the 
  Finite Occurrence Property). 

\item Every $f\in F$ has an associated
    `index' from $N$. 

\item $N$ is linearly ordered by $R$, such that there is a minimal
   element, and each non-maximal element has an immediate successor
   (in particular, if $N$ is infinite, then it contains an infinite
   ascending chain).

\item If $fRg$ for $f,g\in F$, this means that $g$ extends $f$ (as a
   partial bijection), and that the index of $g$ is the successor of
   the index of $f$.

\item The \emph{back-and-forth} properties hold for
    partial isomorphisms whose index is not the maximal element of $N$.

\item Some $f\in F$ has as its index the minimal element of $N$.

\item The submodels defined by $A$ and $B$ disagree on $\phi$.
   (Recall that $\cL$ is closed under the Boolean
   connectives and relativisation by unary predicates).
\end{enumerate}

\item \textbf{Proof of claim:} The first eight properties can already
   be expressed in $FO^3$ by a clever re-use of variables, and the
   ninth property can be expressed in $\cL$ by closure under the
   Boolean connectives and relativisation by unary predicates.

   For instance, the third property can be expressed as the conjunction of all
   $FO^3$-formulas of the following forms, where $S\in REL(\phi)$ is a binary
   relation symbol, and $Q\in REL(\phi)$ is a unary relation symbol:

%   \medskip{\small
%   $\forall xy\Big(Px \land Py \land \exists z(Fz\land Rzx\land Rzy) \to$
%
%   $\big(\exists z(Rxz\land Az\land \exists x(Ryx\land Ax\land Szx)) 
%\leftrightarrow$

%   $\ \ \exists z(Rxz\land Bz\land \exists x(Ryx\land Bx\land Szx))\big)\Big) $
%}
%\hfill and
%\medskip\noindent{\small
%   $\forall x\Big(Px \to \big(\exists z(Rxz\land Az\land Qz) 
%\leftrightarrow \exists z(Rxz\land Bz\land Qz)\big)\Big) $}
\[\eqalign{
  \forall xy\Big(Px \land Py \land \exists z(Fz\land Rzx\land Rzy)
&\to
  \big(\exists z(Rxz\land Az\land \exists x(Ryx\land Ax\land Szx))\cr
&\leftrightarrow\phantom{\big(}
  \exists z(Rxz\land Bz\land \exists x(Ryx\land Bx\land Szx))\big)\Big)\cr
  }
\]
and
\[\forall x\Big(Px \to \big(\exists z(Rxz\land Az\land Qz) 
  \leftrightarrow \exists z(Rxz\land Bz\land Qz)\big)\Big) 
\]
%\smallskip\noindent
Note that we crucially use the fact that the signature
consists of unary and binary relations only. \qed
%\hfill \textbf{End of proof of claim.}
\end{trivlist}

Let $\chi$ be the conjunction of all these $\cL$-sentences. By the
above assumptions, $\chi$ has models in which $N$ has arbitrarily
large finite cardinality (this follows from the existence, for each
$k\in\mathbb{N}$, of models disagreeing on $\phi$ that are potentially
isomorphic up to back-and-forth depth $k$). However, there is no model
of $\chi$ in which $N$ is an infinite set. For, suppose there
were. Let $\theta$ be a formula expressing that a fresh binary
relation symbol $S$ denotes a linear order without maximal element on
the set denoted by $N$, thus forcing $N$ to denote an infinite
set. Then $\chi\land\theta$ would also be satisfiable, hence, by the
L\"owenheim-Skolem property, it would have a countable model.  In
other words, $\chi$ would have a countable model in which $N$ denotes
an infinite set. In countable models, being potentially isomorphic
means being isomorphic; thus, there would be isomorphic models
disagreeing on $\phi$, which contradicts $\cL$'s invariance for
isomorphisms.  

\medskip

In other words, $\chi$ relatively projectively defines
finiteness.
\end{proof}

\begin{thm}\label{thm:strong-Lindstrom-FO}
   An abstract logic extending $FO^3$ is contained in $FO$ iff it
   satisfies both Compactness and the L\"owenheim-Skolem property.
\end{thm}

\begin{proof} If an abstract logic is contained in $FO$, then,
  clearly, it satisfies Compactness and the L\"owenheim-Skolem
  property.  If, on the other hand, an abstract logic $\cL$ extends
  $FO^3$ but is not contained in $FO$, then it must lack either the
  L\"owenheim-Skolem property or Compactness. For, suppose that $\cL$
  satisfies the L\"owenheim-Skolem property and Compactness. Let
  $\psi(N)$ be any $\cL$-sentence projectively defining finiteness
  (cf.~Lemma~\ref{lem:proj-def-finiteness}), and for each
  $k\in\mathbb{N}$, let $\chi_k$ be an $FO^3$-formula expressing that
  there are at least $k$ distinct $N$-elements (by reusing bound
  variables as in $\exists x(Nx \land \exists y(Ny\land Rxy \land
  \exists x(Nx\land Ryx\land \cdots )))$). Then every finite subset of
  $\{\chi_k \mid k\in\mathbb{N}\}\cup\{\psi(N)\}$ has a model while
  the entire set has no model. This contradicts the Compactness
  property. 
\end{proof}

The following results can be proved in a similar fashion:

\begin{thm}\label{thm:strong-Lindstrom-FO-potential-isomorphisms}
   An abstract logic extending $FO^3$ is contained in $FO$ iff it
   satisfies Compactness and invariance for potential isomorphisms.
\end{thm}

\begin{proof}[Proof sketch]
  An analogue of Lemma~\ref{lem:proj-def-finiteness} can be proved where the
  L\"owenheim-Skolem property is replaced by invariance for
  potential isomorphisms. Indeed, the proof is even shorter than the proof of
  Lemma~\ref{lem:proj-def-finiteness} itself, as the last step can 
  be dropped. 
  Given this, we can proceed just as in the proof of Theorem~\ref{thm:strong-Lindstrom-FO}.
\end{proof}

\begin{thm}\label{thm:strong-Lindstrom-FO-recursive-enumerability}
   A concrete abstract logic extending $FO^3$ is contained in $FO$
   iff it satisfies the L\"owenheim-Skolem property and the Finite
   Occurrence Property, and is recursively
   enumerable for validity.
\end{thm}

Here, by ``concrete'' we mean that formulas can be coded as finite
strings over some alphabet, in such a way that negation, conjunction,
and relativisation are computable operations, and there is a
computable translation from $FO^3$-formulas to formulas of the logic.
Theorem~\ref{thm:strong-Lindstrom-FO-recursive-enumerability} follows from
Lemma~\ref{lem:proj-def-finiteness} by closure under relativisation and
the fact that satisfiability of $FO^3$ formulas on finite models is
undecidable \cite{Boerger97:classical}, hence not recursively enumerable.

Note that these results all rely on our restriction to at most
binary relation symbols. In the case with at most $k$-ary relations
($k\geq 1$), analogous results hold for $FO^{k+1}$. For the case of
$k=1$, the argument is slightly different. For the sake of
completeness, we give an outline of it below (similarly, analogues of
Theorems~\ref{thm:strong-Lindstrom-FO-potential-isomorphisms}
and~\ref{thm:strong-Lindstrom-FO-recursive-enumerability} can be shown
for $k=1$). See also the Lindstr\"om theorem for
monadic first-order logic in \cite{Tharp75:which}.

\begin{thm}
  Over unary vocabularies, an abstract logic extending $FO^2$ is
  contained in $FO$ iff it satisfies both Compactness and the
  L\"owenheim-Skolem property. 
\end{thm}

\begin{proof}[Proof sketch]
  If an abstract logic is contained in $FO$, then, clearly, it
  satisfies Compactness and the L\"owenheim-Skolem property.
  Conversely, let $\cL$ be any abstract logic extending $FO^2$ that is
  not contained in $FO$. Let $\phi\in \cL$ be non-first-order, i.e., for
  every $k\geq 0$, there are models $M,N$ disagreeing on $\phi$ such
  that $M$ and $N$ are potentially isomorphic up to depth $k$. 

  By the same argument as in Lemma~\ref{lem:finite-occurrence-FO}, we
  have that $\phi$ depends only on finitely many (unary) predicates
  $P_1, \ldots, P_n$.  Moreover, it can be easily seen that two
  structures are potentially isomorphic up to depth $k$ for $P_1,
  \ldots, P_n$ if and only if for each unary type $\tau$, either
  $\tau$ is realized the same number of times in the two structures,
  or in both structures $\tau$ is realized at least $k$ times. By a
  \emph{unary type} we mean here a function mapping $P_1, \ldots, P_n$
  to truth values. Similarly, two structures are potentially
  isomorphic (for unbounded depth) for $P_1, \ldots, P_n$ if and only
  if for each unary type $\tau$, either $\tau$ is realized the same
  number of times in the two structures, or in both structures $\tau$
  is realized infinitely many times.

  Now, pick fresh unary predicates $Q,\bar{Q}$ and infinitely many fresh
  unary predicates $P_{\tau i}$, for $\tau$ a type and $i\geq 1$.  The
  unary predicates $Q,\bar{Q}$ will be used to define two submodels, which will
  disagree on $\phi$, and the unary predicate $P_{\tau i}$ will be
  used to code a potential isomorphism. As we pointed out already, in
  order to test for the existence of a potential isomorphism it is
  enough to count the number of times each unary type is realized.
  The predicate $P_{\tau i}$, intuitively, will denote the $i$-th element
  of type $\tau$ in one of the two substructures, and they are used to
  compare the cardinalities between the two substructures. More
  specifically, consider the infinite theory $\Sigma$ containing the
  following $FO^2$-sentences:

  \[\begin{array}{l}
  \phi^Q \land \neg\phi^{\bar{Q}} \land  \forall x(\bar{Q}x\leftrightarrow \neg Qx) \\
  \forall x(P_{\tau i}(x) \to \tau(x)) \\ \\
  \forall x (P_{\tau i} (x)\to\neg P_{\tau j}(x)) \text{\quad for $i\neq
    j$}\\ \\
  \forall xy (Qx \land P_{\tau i} x\land Qy \land P_{\tau i}y \to  x=y) \\
  \forall xy (\neg Qx \land P_{\tau i} x\land \neg Qy \land P_{\tau i}y \to
  x=y) \\ \\
  \exists x.(Qx \land P_{\tau (i+1)}x) \to \exists x.(Qx \land x.P_{\tau i}x) \\
  \exists x.(\neg Qx \land P_{\tau (i+1)}x) \to \exists x.(\neg Qx \land
  P_{\tau i}x) \\ \\
  \exists x(Qx \land Px\land \bigwedge_{1 \leq k< i}\neg P_{\tau k}x) \to \exists
  x.(Qx \land P_{\tau i}x) \\
  \exists x(\neg Qx \land Px\land \bigwedge_{1 \leq k <i}\neg P_{\tau k}x) \to \exists
  x.(\neg Qx \land P_{\tau i}x)  \\ \\
  \exists x(Qx\land P_{\tau i}x) \leftrightarrow \exists x(\neg Qx\land
  P_{\tau i}x) 
\end{array}\]
for $\tau$ any type and  $i,j\geq 1$, and where
$\tau(x)$ is shorthand for
$\bigwedge_{\tau(P)=1}P(x)\land \bigwedge_{\tau(P)=0} \neg
P(x)$. The first
sentence expresses that the two substructures disagree on $\phi$. 
The instances of the 2nd -- 9th sentence above force the following:
for all types $\tau$ and natural numbers $i>0$, a substructure (induced
by $Q$ or by $\neg Q$)
contains an element satisfying $P_{\tau i}$ if and only if 
it contains at least $i$ elements of type $\tau$. The instances
of the last sentence above, finally, compare the cardinalities 
between the two structures. It follows that 
the theory $\Sigma$ describes models in which the submodels
induced by $Q$ and $\neg Q$ disagree on $\phi$ and are potentially
isomorphic. 

Incidentally, note that if there are infinitely many elements of type
$\tau$, the theory $\Sigma$ cannot force each to satisfy some
predicate $P_{\tau i}$ (as can be seen by an omitting types argument),
but it does force there to be infinitely many elements satisfying some
predicate $P_{\tau i}$, which suffices for present purposes. 

Since, for every $k\geq 0$, there are models disagreeing on $\phi$
such that the two are potentially isomorphic up to depth $k$, it can
be shown that every finite subset of $\Sigma$ is satisfiable. It
follows that $\cL$ cannot have both Compactness and
L\"owenheim-Skolem, for, the entire theory $\Sigma$ would have a
countable model, and on countable structures, potential isomorphism
implies isomorphism.
\end{proof}

\subsection{First application: Tarski's relation algebra}\label{sec:RA}

Tarski's relation algebra $\mathcal{RA}$ \cite{Tarski87:formalization}
is an algebraic language in which the terms denote binary
relations. It has atomic terms $R, S, \ldots$ ranging over binary
relations (over some domain), constants $\delta$ and $\top$ denoting
the identity relation and the total relation, and operators $\cap, -,
\bullet, \smile$ for taking the intersection, complement,
composition and converse of relations. Thus, the syntax of $\mathcal{RA}$
  is:

\[ \alpha ~::=~ R ~\mid~ \delta ~\mid~ \top ~\mid~ \alpha\cap\beta
~\mid~ \alpha-\beta ~\mid~ \alpha\bullet\beta ~\mid~ \alpha^\smile \] with
$R$ an element from some countably infinite set of variables standing
for binary relations. An interpretation for this language is
a set $X$ together with an assignment of binary relations over $X$ to
the atomic terms. In other words, it is a first-order structure for the
vocabulary that contains the atomic terms as binary relation symbols.  We
write $\alpha\equiv \beta$ if, in each interpretation, $\alpha$ and
$\beta$ denote the same binary relation, and we write
$\alpha\subseteq\beta$ if, in each interpretation, $\alpha$ denotes a
subrelation of the relation denoted by $\beta$.

In this section, we provide a Lindstr\"om-theorem for extensions of
relation algebra.  By an \emph{extended relation algebra} we will mean
any language obtained by extending the syntax of relation algebra with
zero or more additional logical operations, where a \emph{logical
  operation} is any operation that takes as input a fixed finite
number of binary relations $R_1, \ldots, R_n$ (over some set $X$) and
that produces a new binary relation $S$ over the same set.  We also
require logical operations to respect isomorphisms, and to be
\emph{domain independent} in the following sense (familiar from
database theory): the output relation $S$ only depends on part of 
the domain $X$ that participates in at least one pair belonging to an
input relation $R_i$. An example of a domain dependent operation is
the (unary) absolute complement operator, while the (binary) relative
complement operator is domain independent. Another example of a 
domain dependent operator is the zero-ary operator that returns the 
total relation in case the domain $X$ is infinite, and the empty 
relation otherwise. We disallow domain dependent operations, as they
may cause the language to lose closure under relativisation.

\begin{lem}[Closure under relativisation]
  Let $\mathcal{L}$ be any extended relation algebra, let $\alpha$ be
  any expression of $\mathcal{L}$, and let $R$ be any relation
  variable. Then there is an expression of $\mathcal{L}$, which we
  denote by $\alpha^{dom(R)}$ such that in every interpretation $M$,
  the denotation of $\alpha^{dom(R)}$ in $M$ is the denotation of
  $\alpha$ in $M'$, where $M'$ is the submodel of $M$ induced by 
  the domain of $R^M$. 
\end{lem}

\begin{proof}
  First, note that the expression $\top_R := R\bullet \top \bullet
  R^{\smile}$ denote the total relation over the domain of $R$. Now,
  $\alpha^{dom(R)}$ is obtained by uniformly replacing every atomic
  expression $\alpha$ (i.e., the relational variables, as well as
  $\top$ and $\delta$) by $\alpha\cap \top_R$. A straightforward
  formula induction shows that $\alpha^{dom(R)}$ satisfies the
  required conditions. The inductive step for complex expressions uses
  the fact that all operations of $\cL$ are domain independent.
\end{proof}

One example of an extended relation algebra, $\mathcal{RA}_{FO}$, is
the extension of relation algebra with \emph{all first-order definable
   logical operations} (see e.g.~\cite{Venema91:relational}). Another
example is $\mathcal{RAT}$, the extension of relation algebra with the
\emph{transitive closure} operation
\cite{Ng77:relation}.

The Compactness and L\"owenheim-Skolem properties can be defined for
extended relation algebras as usual. For instance, we say that an
extended relation algebra $\mathcal{L}$ has the L\"owenheim-Skolem
property if for every set of $\mathcal{L}$-expressions $\Phi$, if there is
an interpretation under which $\bigcap\Phi$ is a non-empty relation,
then there is such an interpretation over a countable domain.  As is
not hard to see, $\mathcal{RA}$ and $\mathcal{RA}_{FO}$ satisfy both
Compactness and the L\"owenheim-Skolem property, whereas
$\mathcal{RAT}$ satisfies the L\"owenheim-Skolem property but lacks
Compactness.

The following Lindstr\"om-style theorem shows that all extended
relation algebras containing non-first-order definable operations
lack either Compactness or the L\"owenheim-Skolem property.

\begin{thm}\label{thm:Lindstrom-RA}
   Let $\mathcal{L}$ be any extended relation algebra with the
   Compactness and L\"owenheim-Skolem properties. Then every
   logical operation of $\mathcal{L}$ is first-order definable.
\end{thm}

\begin{proof}
   Lemma~\ref{lem:proj-def-finiteness} can be adapted to the relation
   algebra setting, allowing us to show that every extended relation
   algebra containing a non-elementary logical operation and having the
   L\"owenheim-Skolem property can projectively define finiteness,
   and hence lacks Compactness.  We
   will not spell out the details, but merely mention the following
   key points of the proof:
\begin{bDlist}%
\bDitem
   For  every   first-order  sentence  $\phi$   containing  only  three
   variables, in a signature consisting only of binary relations, there
   is a relation algebra expression $\alpha$ such that
   for every model $M$, $M\models\phi$ iff
   $\alpha\equiv\top$ holds in $M$ \cite{Tarski87:formalization}.
\bDitem
   Unary relations can be mimicked by binary relations %
   by systematically intersecting them with the identity relation.
\bDitem
   In this way, every extended relation algebra gives rise to an
   abstract logic extending $FO^3$. Closure under
   relativisation of the logic is guaranteed by the domain
   independence of the logical operations of $\cL$.\qedhere
\end{bDlist}
\end{proof}

In other words, the extension of relation algebra with \emph{all}
elementary operations (a well-known structure in algebraic logic) is 
the greatest extension that satisfies
L\"owenheim-Skolem and Compactness. The same holds if we replace the
L\"owenheim-Skolem property by invariance for potential isomorphisms,
or if we replace Compactness by recursive enumerability.

Theorem~\ref{thm:Lindstrom-RA} nicely complements a known result:
\emph{every extended relation algebra with Craig interpolation can
   define all first-order definable operations}
\cite{tencate2005:interpolation}.  Together, these results show that
$\mathcal{RA}_{FO}$ is the unique (up to expressive equivalence)
extension of $\mathcal{RA}$ satisfying Compactness,
L\"owenheim-Skolem, and Craig Interpolation.

\subsection{Second application: finite variable fragments
}\label{sec:finite-variable}

In this section, we provide Lindstr\"om theorems for the finite
variable fragments $FO^k$ with $k>3$, over vocabularies consisting of
unary and binary relation symbols only.  It is well known that the
finite variable fragments can be characterized as fragments of
first-order logic using potential isomorphisms with a restricted
number of pebbles:

\begin{defi}\textrm{\bf ($k$-pebble potential isomorphisms)}
   A \emph{$k$-pebble potential isomophism} between $M$ and
   $N$ is a non-empty family $F$ of finite partial isomorphisms $f$
   between $M$ and $N$ with $|dom(f)|\leq k$ satisfying the
   following properties: 
\begin{bDlist}
\bDitem \textbf{Closure under restrictions:} for all $f\in F$ and
$f'\subseteq f$, also $f'\in F$. 
\bDitem \textbf{Forth:} for all $\{(w_1,v_1),\ldots, 
(w_n,v_n)\}\in F$ with
   $n<k$, and $w\in M$, there is an $v\in N$ such that
    $\{(w_1,v_1), \ldots, (w_n,v_n),(w,v)\}\in F$
\bDitem \textbf{Back:} for all $\{(w_1,v_1),\ldots, 
(w_n,v_n)\}\in F$ with
   $n<k$, and $v\in N$, there is an $w\in M$ such that
    $\{(w_1,v_1), \ldots, (w_n,v_n),(w,v)\}\in F$
\end{bDlist}
\end{defi}

\begin{fact} \label{fact:finite-variable}
   $FO^k$ is (up to logical equivalence) the fragment of $FO$ invariant
   for $k$-pebble potential isomorphisms ($k\geq 1$).
\end{fact}

For a proof of Fact~\ref{fact:finite-variable} see for instance
\cite{Andreka98:modal}.

Using Lemma~\ref{lem:proj-def-finiteness}, we can turn this into the
following Lindstr\"om characterization (remember that we only consider
unary and binary relation symbols):

\begin{thm}[Lindstr\"om theorem for $FO^k$]\label{thm:Lindstrom-FOk}
   Let $k\geq 3$.  An abstract logic extending $FO^k$ satisfies
   Compactness and invariance for $k$-pebble potential isomorphisms iff
   it is no more expressive than $FO^k$.
\end{thm}

\begin{proof}
   Consider any abstract logic $\cL$ extending $FO^3$ that has Compactness
   and is invariant for $k$-pebble potential isomorphisms. Then in particular
   it is invariant for potential isomorphisms, and therefore by
   Theorem~\ref{thm:strong-Lindstrom-FO-potential-isomorphisms} it must
   be contained in $FO$. But then, by Fact~\ref{fact:finite-variable},
   it must be contained in $FO^k$.
\end{proof}

Theorem~\ref{thm:Lindstrom-FOk} can also be seen as a strengthening of
the classical preservation result in Fact~\ref{fact:finite-variable}
(for $k\geq 3$, and on binary vocabularies). Indeed, it implies (and in
fact is equivalent to, as we will explain) the following
``generalized preservation theorem'': 

\begin{cor} \label{cor:gen-pres-FOk}
   Let $k\geq 3$. Let $\cL$ be any abstract logic extending $FO^k$ with
   Compactness. Then $FO^k$ is the fragment of $\cL$ invariant for
   $k$-pebble potential isomorphisms (up to logical equivalence). In
   particular, $FO^k$ is the fragment of $FO$ invariant for
   $k$-pebble potential isomorphisms.
\end{cor}

\begin{proof}
   Let $\cL$ be any compact abstract logic extending $FO^k$, and let
   $\cL'$ be the fragment of $\cL$ invariant for $k$-pebble potential
   isomorphisms. Then $\cL'$ satisfies all the requirements of abstract
   logics. For instance, it is closed under relativisation: if
   $\phi\in \cL$ is invariant for $k$-pebble potential isomorphisms,
   then so is its relativisation by a unary predicate.
   Thus, $\cL'$ is an abstract logic extending
   $FO^k$ that is compact and invariant for $k$-pebble potential
   isomorphisms. Hence, by Theorem~\ref{thm:Lindstrom-FOk}, it is
   contained in $FO^k$.
\end{proof}

In fact, Corollary~\ref{cor:gen-pres-FOk} also implies
Theorem~\ref{thm:Lindstrom-FOk}: let $\mathcal{L}$ be any logic
extending $FO^k$ satisfying Compactness and invariance for $k$-pebble
potential isomorphisms. By Corollary~\ref{cor:gen-pres-FOk}, $FO^k$ is
the the fragment of $\cL$ invariant for $k$-pebble potential
isomorphisms. But since $\cL$ itself is invariant for $k$-pebble
potential isomorphisms, this means that $\cL$ and $FO^k$ coincide in
terms of expressive power. This shows that
Corollary~\ref{cor:gen-pres-FOk} and Theorem~\ref{thm:Lindstrom-FOk}
are really equivalent to each other: Lindstr\"om theorems and 
preservation theorems sometimes amount to the same thing.

\section{From modal logic upwards}

The approach of generalizing the classic Lindstr\"om theorem only got
us so far. It enabled us to characterize $FO^k$ for $k\geq 3$ but is
unlikely to reach much further down, since coding power then falls 
under the minimum needed to describe partial isomorphisms and their 
extension properties.  Thus, we will now take a different
approach, by considering the modal Lindstr\"om theorem, and trying to
generalize it to richer languages. In particular, we obtain
two new Lindstr\"om theorems for the graded modal logic.

\subsection{The modal Lindstr\"om theorem revisited}\label{sec:Lindstrom-ML}

We recall the proof of the modal Lindstr\"om theorem of
\cite{benthem07:lindstrom} (which improves on an earlier result in
\cite{Rijke95:Lindstroem}).  First, we need to define ``abstract modal
logics''.  As before, we assume a fixed vocabulary, now
consisting of a single binary relation symbol $R$ and a countably
infinite set of unary relation symbols, also called \emph{proposition
   letters}. Structures for this vocabulary are usually called
\emph{Kripke models} (the restriction to a single binary relation
symbol is not essential but is convenient for presentation).
We associate to each formula a class of pairs $(M,w)$, where $M$ is a
Kripke model and $w$ is an element of the domain of $M$. This is
because modal formulas are always evaluated \emph{at a point in a
   model}. We call such pairs $(M,w)$ \emph{pointed Kripke
   models}.  Thus, an \emph{abstract modal logic} is a pair
$\cL=(Fml_{\cL},\models)$, where $Fml_{\cL}$ is the set of formulas of
$\mathcal{L}$ and $\models$ is a binary relation between
$\mathcal{L}$-formulas and pointed Kripke models. As before, when no
confusion arises, we will write $\cL$ for $Fml_{\cL}$. Also, as before,
we assume that $\cL$-formulas are invariant for isomorphisms, and that
$\cL$ is closed under the Boolean operations, renaming, and
relativisation by unary predicates. \footnote{So far, there is nothing particularly 
`modal' about these systems, and we might also speak of ``abstract local logics'', 
or some other appealing terminology. In particular, we do \emph{not}  
replace isomorphism invariance by `modal' invariance for bisimulations
in the definition of these logics. The power of the latter 
condition is precisely one of the things we want to scrutinize in what follows.}

Examples of abstract modal logics include \emph{basic modal logic},
its extension with counting modalities called \emph{graded modal
   logic} ($GML$), first-order logic (by which we mean the collection of
all first-order formulas in one free variable, over the appropriate
signature), and the modal $\mu$-calculus. For the syntax and semantics
of basic modal logic, see any modal logic
textbook (e.g.,~\cite{Blackburn01:modal}).

\medskip

The modal Lindstr\"om theorem characterizes basic modal logic
in terms of Compactness and invariance for \emph{bisimulations}. A
bisimulation between Kripke models $M$ and $N$ is a binary relation
$Z$ between the domains of $M$ and $N$ satisfying the following three
conditions:

\begin{bDlist}
\bDitem \textbf{Atomic harmony:} if $wZv$ then $w$ and $v$ agree on all
       proposition letters (unary predicates).
\bDitem \textbf{Zig:} if $wZv$ and $wR^M w'$, there is a $v'$ such 
that $vR^N v'$ and $w'Zv'$.
\bDitem \textbf{Zag:} if $wZv$ and $vR^N v'$, there is a $w'$ such 
that $wR^M w'$ and $w'Zv'$.
\end{bDlist}
Two pointed Kripke models $(M,w)$ and $(N,v)$ are %
\emph{bisimilar} if there is a bisimulation $Z$ between $M$ and $N$
with $wZv$. A formula is
\emph{bisimulation invariant} if it does not distinguish
bisimilar pointed Kripke models, and an abstract modal logic is
bisimulation invariant if all its formulas are.

Given a pointed Kripke model $(M,w)$, we denote by $M_w$ the submodel
of $M$ containing all points that are reachable in finitely many steps
from $w$ along the binary relation. Likewise, for $k\in\mathbb{N}$,
$M_w^k$ is the submodel of $M$ containing all points reachable from
$w$ in at most $k$ steps along the binary relation.  We say that a
formula $\phi$ is \emph{invariant for generated submodels} if, for all
models $M$ with nodes $w$, $(M,w)\models\phi$ iff
$(M_w,w)\models\phi$. We say that $\phi$ \emph{has the finite depth
   property} if there is a $k\in \mathbb{N}$ such that
$(M,w)\models\phi$ iff $(M_w^k,w)\models\phi$, for all models $M$ with
nodes $w$. Clearly, the latter implies the former. Also, bisimulation
invariance implies invariance for generated submodels, because the
natural inclusion map is a bisimulation.
An abstract modal
logic $\cL$ is %
invariant for generated submodels (or, %
has the finite depth property), if every $\phi\in\cL$ is invariant
for generated submodels (respectively, has the finite depth property).

\medskip

We are now ready to proceed with the proof of the modal Lindstr\"om
theorem, Theorem~\ref{thm:Lindstrom-ML} below. We first prove a finite
occurrence property for compact extensions of basic modal logic that
satisfy invariance for bisimulation (in fact, invariance for 
generated submodels will do).

\begin{lem}[Finite occurrence property]\label{lem:ml-finite-occurrence}
   Let $\cL$ be any abstract modal logic extending basic  modal
   logic that is compact and invariant for generated submodels.
   Then for each $\phi\in\cL$, there is a finite set of proposition
   letters $PROP(\phi)$ such that the truth of $\phi$ in any pointed
   Kripke model is independent of the denotation of proposition letters
   outside $PROP(\phi)$.
\end{lem}

\begin{proof}
   Like that of Lemma~\ref{lem:finite-occurrence-FO}.  Since the set of
   proposition letters (unary predicates) is infinite, we can find
   renamings $\rho_1, \rho_2$ for them whose range is disjoint. Now,
   take any $\phi\in\cL$, and let $\phi_1, \phi_2$ be its renamings
   according to $\rho_1$ and $\rho_2$. Let $\Sigma =
   \{\Box^n(\rho_1(p)\leftrightarrow\rho_2(p))\mid n\in\mathbb{N}
   \text{ and $p$ a proposition letter}\}$, where $\Box^n$ stands for a
   sequence of $n$ boxes. It follows from generated submodel invariance
   that $\Sigma\models \phi_1\leftrightarrow\phi_2$,  hence, by
   Compactness, a finite subset $\Sigma'\subseteq\Sigma$ implies
   $\phi_1\leftrightarrow\phi_2$. We can pick for
   $PROP(\phi)$ the set of proposition letters occurring in
   $\Sigma'$.
\end{proof}

Our crucial observation in this Lindstr\"om argument is the following:

\begin{lem}[\cite{benthem07:lindstrom}]\label{lem:ml-finite-depth}
   Let $\cL$ be any abstract modal logic extending basic modal logic that is
   compact and invariant for generated submodels. Then $\cL$ has
   the finite depth property. 
\end{lem}

\begin{proof}
   Let $\phi\in \cL$, with $p$ a proposition letter not
   occurring in $\phi$, and let $\phi^p$ be the relativisation of
   $\phi$ by $p$.  By the generated submodel-invariance of $\cL$, $\{p,
   \Box p, \Box\Box p, \ldots\} \models \phi\leftrightarrow\phi^p$.  By
   the compactness of $\cL$, there is an $n\in\mathbb{N}$ such that
   $\{p, \Box p, \ldots, \Box^n p\} \models \phi\leftrightarrow
   \phi^p$.  But this expresses exactly that $\phi$ has the finite
   depth property, for depth $n$.
\end{proof}

\begin{thm}[\cite{benthem07:lindstrom}]\label{thm:Lindstrom-ML}
   An abstract modal logic extending basic modal
   logic satisfies Compactness and bisimulation invariance iff
   it is no more expressive than basic modal logic.
\end{thm}

\begin{proof}
   Let $\cL$ be any abstract modal logic extending basic modal
   logic and satisfying Compactness and bisimulation invariance.
   Since bisimulation invariance implies invariance for generated
   submodels, $\cL$ is also invariant
   for generated submodels. Then, by
   Lemma~\ref{lem:ml-finite-occurrence} and
   Lemma~\ref{lem:ml-finite-depth}, it follows that $\cL$ has the 
finite occurrence
   property and the finite depth property.  Next, we use the following
   well known fact \cite{Blackburn01:modal}:
   \begin{quote}%
   Assuming a finite vocabulary, every bisimulation-invariant class
   of pointed models with the finite depth property is definable by
   a formula of basic modal logic.
   \end{quote}
We conclude that $\cL$ is contained in basic modal logic.
\end{proof}

Theorem~\ref{thm:Lindstrom-ML} can be seen as a strengthening of van Benthem's
more familiar characterization of basic modal logic as the bisimulation
invariant fragment of first-order logic. Indeed, it implies the 
latter theorem (as observed in \cite{benthem07:lindstrom}), but 
something even stronger holds, viz. the
following ``generalized preservation theorem'':

\begin{cor}\label{cor:bisimulation-preservation}
   Let $\cL$ be any abstract modal logic extending basic modal
   logic that has Compactness. Then basic modal logic is the
   bisimulation invariant fragment of $\cL$ (up to logical
   equivalence). In particular, basic modal logic is the
   bisimulation invariant fragment of first-order logic.
\end{cor}

\begin{proof}
   Let $\cL'$ be the bisimulation invariant fragment of $\cL$.  Then
   $\cL'$ satisfies all criteria for being an abstract modal logic.  For
   instance it is closed under relativisation:  whenever
   $\phi\in\cL$ is invariant for bisimulations then the relativisation
   of $\phi$ by a unary predicate is also invariant for bisimulations.
   Likewise for the Boolean connectives.

   Hence, $\cL'$ is an abstract modal logic extending basic modal logic and
   it is bisimulation invariant and Compact. Hence, it is no more
   expressive than basic modal logic.
\end{proof}

Corollary~\ref{cor:bisimulation-preservation} strengthens
the traditional bisimulation preservation theorem, as there are
compact extensions of basic modal logic not contained
in first-order logic. Indeed:

\begin{thm}\label{thm:well-behaved-extension}
   There is an abstract modal logic extending basic modal logic that is
   not contained in first-order logic, but still satisfies Compactness,
   the L\"owenheim-Skolem property, invariance for potential
   isomorphisms, and Craig interpolation. Moreover, it has a finite
   axiomatization and a \textsc{PSpace}-complete satisfiability
   problem.
\end{thm}

The proof is given in the next section. Note that
Theorem~\ref{thm:well-behaved-extension} is also interesting for another
reason: it shows that
Theorem~\ref{thm:strong-Lindstrom-FO} no longer holds when $FO^3$ is
replaced by basic modal logic.

By the same argument as before in the case of finite variable logics,
Corollary~\ref{cor:bisimulation-preservation} not only follows from
Theorem~\ref{thm:Lindstrom-ML} but also implies it, and hence the two
are equivalent. Our results resolve a question posed in
\cite{benthem07:lindstrom} concerning the relationship between the
modal Lindstr\"om theorem and the `classic' bisimulation preservation
theorem.

\subsection{A well-behaved non-elementary extension of modal
   logic}\label{sec:well-behaved-extension}

In this section, we prove Theorem~\ref{thm:well-behaved-extension}:
we introduce an extension of modal logic,
$ML^{\bullet}$, that is non-elementary but at the same time is very
well behaved (e.g., it has the Compactness and L\"owenheim-Skolem properties).
  $ML^{\bullet}$ extends basic modal logic with an
extra operator $\bullet$. Thus, the formulas of $ML^{\bullet}$ are
given by:
\[ \phi ::= p \mid \neg\phi \mid \phi\land\psi \mid \Diamond\phi \mid
     \bullet\phi \]
     The semantics of the newly added operator is as follows:
     \emph{$M,w\models\bullet\phi$ iff $w$ has infinitely many
       reflexive successors satisfying $\phi$} (a reflexive node is
     one related to itself).

     The reader may verify that $\bullet$ behaves as any normal modal
     operator: $\bullet(\phi\lor\psi)$ is equivalent to
     $\bullet\phi\lor\bullet\psi$, and $\bullet\bot$ is equivalent to
     $\bot$. The dual of $\bullet$ is denoted
     by $\bar{\bullet}$.

\begin{prop}
   $ML^{\bullet}$ is a non-elementary abstract modal logic extending
   basic modal logic.
\end{prop}

\begin{proof}
   This is an easy verification: e.g., the language is
   closed under relativisation.
\end{proof}

\begin{prop}
   $ML^{\bullet}$ has the L\"owenheim-Skolem property and is invariant 
for potential isomorphisms.
\end{prop}

\begin{proof}
By the containment of $ML^{\bullet}$ in
$L_{{\omega_1}\omega}$.
\end{proof}

Before we proceed to prove the remaining properties of $ML^{\bullet}$,
we will first provide an alternative semantics for the logic.  Let $K$
be the class of all bi-modal Kripke frames
$\mathfrak{F}=(W,R_\Diamond,R_\bullet)$ such that $R_\bullet\subseteq
R_\Diamond$ and for all $(w,v)\in R_\bullet$, $(v,v)\in R_\Diamond$.
Kripke models based on such frames can be thought of as quasi-models
for $ML^{\bullet}$. The next lemma shows that they are adequate as
such.

\begin{lem}\label{lem:alternative-semantics}
   For all $\phi\in ML^{\bullet}$, $\phi$ is satisfiable according to
   the intended semantics iff $\phi$ is satisfiable on the class $K$.
\end{lem}

\begin{proof}
   In one direction, suppose that $M,w\models\phi$ according to the
   intended semantics, where $M=(W,R,V)$ is a uni-modal Kripke
   model. Define an equivalence relation $\sim$ on $W$ by letting
   $v\sim v'$ iff $v$ and $v'$ assign the same truth value to all
   subformulas of $\phi$.  Let $R_\Diamond := R$ and $R_\bullet :=
   \{(v,v')\in R\mid (v',v')\in R$ and there are infinitely many
   $v''\sim v'$ such that $(v,v'')\in R$ and $(v'',v'')\in R \}$. Then
   the underlying Kripke frame of $M'$ belongs to $K$, and a
   straightforward inductive argument shows that, for all subformulas
   $\psi$ of $\phi$, and for all $v\in W$, $M,v\models\psi$ according
   to the intended semantics iff $M',v\models\psi$. In particular,
   $M',w\models\phi$. Incidentally, it is important for the above
   argument that $\sim$ be an equivalence relation of finite index, so
   that the pigeon hole principle can be applied in the inductive 
clause for the new modality.

   Conversely, let $M,w\models\phi$, where
   $M=(W,R_\Diamond,R_\bullet,V)$ is a bi-modal Kripke model based on a
   frame in $K$. We will construct a uni-modal model $M'$ with a node
   $w'$ such that $M',w'\models\phi$ according to the intended semantics.
   Roughly, the construction of $M'$ is based on the following ideas:
   (i) make sure that all $R_\Diamond\setminus R_\bullet$-successors of
   a node are irreflexive, by \emph{unraveling}, and (ii) taking infinitely
   many copies of each $R_\bullet$ successor of each node.

   More precisely, as the domain of $M'$, we choose $W\times\mathbb{N}$.
   The accessibility relation $R$ is the set of all pairs
   $((v,n),(v',m))$ satisfying one of the following conditions:
\begin{enumerate}[$-$]
\item $m\leq 1$, $m\neq n$, and $(v,v')\in R_\Diamond$; or
\item $m\geq 2$, and $(v,v')\in R_\bullet$.
\end{enumerate}
Observe that, for all $v\in W$, $(v,0)$ and $(v,1)$ are
$R$-irreflexive by  construction, whereas $(v,n)$ for $n\geq 2$
is $R$-reflexive or $R$-irreflexive, depending on $v$.

Finally, the valuation function $V'$ for the atomic propositions is
defined as usual, by letting $V'(p) = \{(v,n)\mid v\in V(p)\}$.  Let
$M'=(W\times\mathbb{N},R,V')$.

A straightforward inductive argument now shows that, for all formulas
$\psi$, nodes $v\in W$ and natural numbers $n$, $M,v\models\psi$ iff
$M',(v,n)\models\psi$ according to the intended semantics.
In particular, we see that
$M',(w,0)\models\phi$.
\end{proof}

In the remainder of this subsection, we use this alternative semantics.

\begin{prop}\label{prop:completeness}
   A complete axiomatization for $ML^{\bullet}$ can be obtained by extending
   any standard axiomatization for basic (bi-)modal logic with the
   following axioms:
   \[\bullet\phi\to\Diamond\phi \qquad \text{ and } \qquad
     \bar{\bullet}(\phi\to\Diamond\phi)\]
   This axiomatization is even strongly complete: each consistent
   set of formulas is jointly satisfiable.
\end{prop}

\begin{proof}
   Both axioms are Sahlqvist formulas, and together, they define the 
frame class $K$. It follows by the Sahlqvist completeness
   theorem that the axiomatization is strongly complete for $K$: every
   consistent set of formulas is jointly satisfiable in a Kripke model
   based on a frame in $K$.

   It follows by Lemma~\ref{lem:alternative-semantics} that the
   axiomatization is also sound and strongly complete with respect to
   the intended semantics of $ML^{\bullet}$ (the proof of
   Lemma~\ref{lem:alternative-semantics} actually shows that every set
   of formulas jointly satisfiable on $K$ is jointly satisfiable
   according to the intended semantics).
\end{proof}

It follows that $ML^{\bullet}$ also satisfies Compactness. We add a 
few further interesting properties.

\begin{lem} The satisfiability problem for $ML^{\bullet}$ is
   \textsc{PSpace}-complete.
\end{lem}

\begin{proof}
   The lower bound follows from the \text{PSpace}-hardness of basic
   modal logic. As for the upper bound, the standard
   \text{PSpace}-algorithm for basic modal logic due to Ladner can
   be extended in a straightforward manner to $ML^{\bullet}$ (using the
   alternative semantics provided by
   Lemma~\ref{lem:alternative-semantics}).

   Alternatively, one can also perform a direct reduction:
   for any $ML^{\bullet}$ formula $\phi$, let
   $\phi^*$ be the uni-modal formula obtained by replacing each
   subformula of the form $\bullet\psi$ by $\Diamond(r\land\psi)$, for
   $r$ a (fixed) fresh proposition letter.  Let $Sub(\phi)$ be the set
   of subformulas of $\phi$, and let $depth(\phi)$ be the modal nesting
   depth of $\phi$.

   It can be shown that, for any $\phi\in ML^{\bullet}$, $\phi$ is 
satisfiable on $K$ iff
   \[\phi^*\land\mathop{\bigwedge_{k\leq depth(\phi)}}_{\psi\in Sub(\phi)}
   \Box^k(r\land\psi\to\Diamond\psi)\] is a satisfiable basic modal formula.
   Roughly, the idea is that the additional conjuncts in the above
   formula guarantee that we can make every $r$-node reflexive
   in the model without affecting the truth of $\phi$. Together with
   Lemma~\ref{lem:alternative-semantics} this gives us a polynomial time
   reduction from $ML^{\bullet}$-satisfiability to satisfiability of
   basic modal formulas.
\end{proof}

\begin{prop}
   $ML^{\bullet}$ has the Craig Interpolation property: for all
   $ML^{\bullet}$-formulas $\phi,\psi$, if $\models\phi\to\psi$ 
  then there is an $ML^{\bullet}$-formula $\chi$ such that 
  $\models\phi\to\chi$, $\models\chi\to\psi$, and all proposition 
  letters occurring in $\chi$ occur both in $\phi$ and in $\psi$. 
\end{prop}

\begin{proof}
  This follows from a result in \cite{Marx97:multi}, according to
  which every multi-modal logic axiomatized by Sahlqvist formulas with
  first-order universal Horn correspondents has Craig interpolation in
  the sense described above (note that we do not require the modal
  operators occurring in $\chi$ to occur both in $\phi$ and in
  $\psi$). The two axioms from Proposition~\ref{prop:completeness} are
  precisely of this form. 
\end{proof}

Note that $ML^\bullet$ is \emph{not} invariant for bisimulations, and
it does \emph{not} have the finite model property. 
It would be of interest to characterize the extensions of basic 
modal logic satisfying Compactness and the Finite Model Property. 
Perhaps this could still lead to some sort of modal counterpart 
to the Compactness-L\"owenheim-Skolem version of the 
Lindstr\"om theorem for first-order logic.

\subsection{Graded modal logic}\label{sec:$GML$-arbitrary}

Graded modal logic ($GML$) extends basic modal logic
with counting modalities: for each formula $\phi$ and natural number
$k$, $\Diamond_k\phi$ is admitted as a formula, and it says that
at least $k$ successors of the current node satisfy $\phi$.

\newcommand{\rt}{\mathsf{root}}
\newcommand{\unr}{\mathsf{unr}}

$GML$-formulas are in general not invariant for
bisimulations. Still, an important weaker invariance property does
hold: $GML$ formulas are invariant for \emph{tree unraveling}. A
\emph{tree model} is a Kripke model whose underlying frame
is a tree (in the graph theoretic sense, possibly infinite, and with
a unique root). We will denote tree models by $T, T', \ldots$, and we
will use $\rt(T)$ to denote the root of the tree model $T$.  Every
pointed Kripke model $(M,w)$ can be unraveled into a tree model, by
the following standard construction:

\begin{defi}[Tree unraveling]
Given a Kripke model $M=(W,R,V)$ and $w\in W$, the tree unraveling
   $\unr(M,w)$ is defined as $(W',R',V')$, where
\begin{bDlist}
\bDitem $W'$ consists of all finite paths $\langle w_1, w_2, \ldots, 
w_n\rangle$
   satisfying $w_1=w$ and $w_i R w_{i+1}$.
\bDitem $R'$ contains all pairs of sequences of the form $(\langle w_1, \ldots,
   w_n\rangle, \langle w_1, \ldots, w_n, w_{n+1}\rangle)\in W'\times W'$
\bDitem $\langle w_1, \ldots, w_n\rangle\in V'(p)$ iff  $w_n\in V(p)$.
\end{bDlist}
\end{defi}

It is easily seen that, for any pointed Kripke model $(M,w)$, $\unr(M,w)$ is
indeed a tree model, and that $\langle w \rangle$ is its root.
$GML$-formulas are invariant for this operation:

\begin{fact}[$GML$ is invariant for tree unravelings]
   For all pointed Kripke models $(M,w)$ and $GML$-formulas $\phi$,
   $M,w\models\phi$ iff $\unr(M,w),\langle w\rangle\models\phi$.
\end{fact}

The proof is a straightforward formula induction.

We will prove two Lindstr\"om theorem for $GML$. The first characterizes
$GML$ on arbitrary structures in terms of Compactness, the
L\"owenheim-Skolem property and invariance for tree unravelings. It
can be seen as a natural generalization of
Theorem~\ref{thm:Lindstrom-ML}.  The second theorem, which
will be proved in the following section, considers $GML$ as a language
for describing nodes in \emph{tree models}, and it characterizes $GML$ as
being maximal with respect to Compactness and the L\"owenheim-Skolem
property on such structures.

Theorem~\ref{thm:Lindstrom-ML} characterized modal logic
in terms of Compactness and bisimulation invariance. One might wonder
if, likewise, $GML$ can be characterized by Compactness and
invariance for tree unraveling. The answer is negative: the extension
of $GML$ with the modal operator $\Diamond_{\aleph_1}$ (``uncountably
many successors \ldots'') still has these properties (even
first-order logic extended with the ``\emph{uncountably many}''
quantifier has (countable) Compactness \cite{Fuhrken64:Skolem}).  Instead, we
prove the following:

\begin{thm}
   \label{thm:lindstrom-$GML$}
   An abstract modal logic extending $GML$
   satisfies invariance for tree unravelings,
   Compactness, and the L\"owenheim-Skolem property iff
   it is no more expressive than $GML$.
\end{thm}

As in the case of modal logic, we obtain the following
``generalized preservation theorem'' as a corollary (the proof is
analogous to that of Corollary~\ref{cor:bisimulation-preservation}):

\begin{cor}
   Let $\cL$ be any abstract modal logic extending $GML$ and
   satisfying Compactness and the L\"owenheim-Skolem property.
   Then $GML$ is the tree unraveling invariant fragment of $\cL$
   (up to logical equivalence). In particular, $GML$ is the tree 
unraveling invariant fragment of $FO$.
\end{cor}

The rest of this section is dedicated to the proof of
Theorem~\ref{thm:lindstrom-$GML$}.  Two easily established facts about
$GML$ will be used in the proof:

\begin{fact}[$GML$ has the finite tree model property]\label{prop:fmp}
Every satisfiable $GML$ formula is satisfied at the root of some
finite tree model.
\end{fact}

\begin{fact} \textrm{\bf($GML$ can describe finite tree models up to 
isomorphism)} \quad
   Assuming a finite vocabulary,
   for every finite tree model $T$, there is a $GML$-formula
   $\psi_{T}$ such that for every tree model $T'$,
   $(T',\rt(T'))\models\psi_{T}$ iff $T'\cong T$.
\end{fact}

Fix an abstract modal logic $\cL$ extending $GML$ and satisfying
Compactness and L\"owen\-heim-Skolem, as well as invariance for tree
unravelings.  By Lemmas~\ref{lem:ml-finite-occurrence}
and~\ref{lem:ml-finite-depth}, $\cL$ has the finite occurrence
property and the finite depth property (note that invariance for tree
unraveling implies invariance for generated submodels).
The following Lemma shows a kind of ``finite
width property'':

\begin{lem} \textrm{\bf (Can only count successors up to a finite
    number)} \quad \label{lem:finite-width} Let $\cL$ be any abstract modal
  logic extending $GML$ and satisfying Compactness,
  L\"owenheim-Skolem, and invariance for tree unravelings. Then for
  each formula $\phi\in \cL$ and finite tree model $T$, there is a
  natural number $k$ such that ``$\phi$ can only count $T$-successors
  up to $k$'': whenever a tree model contains a node $v$ that has $k$
  successor subtrees isomorphic to $T$, then adding more copies of $T$
  will not affect the truth value of $\phi$ at the root.
\end{lem}

\begin{proof}
   Since $T$ is a finite tree model, it can be described up to
   isomorphism by a single $GML$-formula $\psi_{T}$.  Let $\Sigma$ be
   the following set of formulas, where $p$ is a proposition letter not
   occurring in $\phi$, and $\Box^n$ stands for a sequence of $n$
   boxes:
%
%   \medskip\noindent
%   \begin{tabular}{p{.05\linewidth}@{}p{.9\linewidth}}
%     & $\{p,~ \Box^n(\neg p \to \Box\neg p) \mid n\in\mathbb{N}\}$ \\
%     & ``$p$ defines an initial subtree''
%   \end{tabular}
%
%   \noindent\begin{tabular}{p{.05\linewidth}@{}p{.9\linewidth}}
%    $\cup$ & $\{ \Box^n(p\to \Box(\neg p\to \psi_{T})) \mid n\in\mathbb{N}\}$ \\
%     & ``the root of every $\neg p$-subtree satisfies $\psi_{T}$''
%   \end{tabular}
%
%   \noindent\begin{tabular}{p{.05\linewidth}@{}p{.9\linewidth}}
%    $\cup$ & $\{ \Box^n(p\land \Diamond\neg p \to \Diamond_m (p\land 
%\psi_{T})) \mid n,m\in\mathbb{N}\}$ \\
%     &  ``every $p$-node with a $\neg p$-successor has infinitely many 
%$p$-successors satisfying $\psi_{T}$''
%   \end{tabular}
\[\eqalign{
   &\{p,~ \Box^n(\neg p \to \Box\neg p) \mid n\in\mathbb{N}\}\cr
   &\raise 4 pt\hbox{``$p$ defines an initial subtree''}\cr
  \cup{\enspace}
   &\{ \Box^n(p\to \Box(\neg p\to \psi_{T})) \mid n\in\mathbb{N}\}\cr
   &\raise 4 pt\hbox{``the root of every $\neg p$-subtree satisfies $\psi_{T}$''}\cr
  \cup{\enspace}
   &\{ \Box^n(p\land \Diamond\neg p \to \Diamond_m (p\land 
    \psi_{T})) \mid n,m\in\mathbb{N}\}\cr
   &\raise 4 pt\hbox{``every $p$-node with a $\neg p$-successor has infinitely many 
    $p$-successors satisfying $\psi_{T}$''}\cr
  }
\]
   \begin{figure}
   \begin{center}
   \includegraphics[scale=0.9]{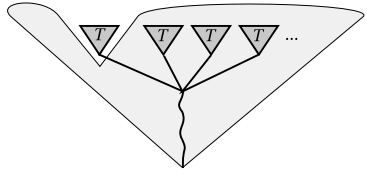}
   \end{center}
   \caption{Model from the proof of Lemma~\ref{lem:finite-width}}
   \label{fig:cutting}
   \end{figure}
   \noindent
   Whenever a \emph{countable tree model} satisfies $\Sigma$ at the
   root,  the submodel defined by $p$ is isomorphic to the whole
   model --- isomorphic in the language without $p$, to be precise (see
   Figure~\ref{fig:cutting}). Since $\cL$ satisfies the
   L\"owenheim-Skolem property, invariance for tree unraveling and
   invariance for isomorphisms, we can conclude that
   $\Sigma\models\phi\leftrightarrow\phi^p$.
   But then, by compactness, there is a $k\in\mathbb{N}$ such that
   $\Sigma_k\models\phi\leftrightarrow\phi^p$, where $\Sigma_k$
   is the following subset of $\Sigma$:
%   \medskip\noindent{\begin{tabular}{p{.95\linewidth}}
%   $\{p,~ \Box^n(\neg p \to \Box\neg p) \mid n\in\mathbb{N}\} {~\cup}$   \\[1mm]
%   $\{ \Box^n(p\to \Box(\neg p \to \psi_{T})) \mid n\in\mathbb{N}\} 
%{~\cup}$  \\[1mm]
%   $\{ \Box^n(p\land \Diamond\neg p \to \Diamond_m (p\land \psi_{T})) 
%\mid n,m\in\mathbb{N}\text{ \emph{with $m\leq k$}}\}$
%   \end{tabular}}
%   \smallskip\par\noindent
\[\eqalign{
  &\{\,p,~ \Box^n(\neg p \to \Box\neg p) \mid n\in\mathbb{N}\,\}\cr
\cup{\enspace}
  &\{\,\Box^n(p\to \Box(\neg p \to \psi_{T})) \mid n\in\mathbb{N}\,\}\cr
\cup{\enspace}
  &\{\,\Box^n(p\land \Diamond\neg p \to \Diamond_m (p\land \psi_{T})) 
    \mid n,m\in\mathbb{N}\text{ \emph{with $m\leq k$}}\, \}\cr
  }
\]
   This shows that the Lemma holds.
\end{proof}

\begin{proof}[Proof of Theorem~\ref{thm:lindstrom-$GML$}]
  Let $\cL$ be any abstract modal logic extending $GML$, satisfying
  Compactness, the L\"owenheim-Skolem property, and tree unraveling
  invariance. Observe that $\cL$ still satisfies Compactness and the
  L\"owenheim-Skolem property if we restrict attention to trees (the
  tree unraveling of a countable model is countable).  Also, by
  Lemmas~\ref{lem:ml-finite-occurrence} and~\ref{lem:ml-finite-depth},
  $\cL$ has the finite occurrence property and the finite depth
  property (note that invariance for tree unraveling implies
  invariance for generated submodels).

  Consider any formula $\phi\in \cL$. We will construct a set of
  equivalence relations $\sim^\phi_i$ for tree models (with $i\geq
  0$), satisfying the following two properties:
\begin{enumerate}[(1)]%\setlength{\itemsep}{0mm}
\item $T\sim^\phi_i T'$ implies that the truth value of $\phi$ at the
   root of a tree model is not affected if subtrees isomorphic to $T$
   at depth $i$ are replaced by copies of $T'$ (or vice versa).
\item Each $\sim^\phi_i$ has only finitely many equivalence classes,
   and each is definable by a $GML$-formula.
\end{enumerate}

This then implies that $\phi$ is equivalent to a $GML$ formula (take the
disjunction of the $GML$-formulas defining the $\sim^\phi_0$-equivalence
classes that satisfy $\phi$).

The claim holds trivially for $i>
depth(\phi)$. Next, assume that the claim holds for $i+1$. We will
show that it also holds for $i$. Let $K_1, \ldots, K_n$ be the
(finitely many) $\sim^\phi_{(i+1)}$-equivalence classes, and for each
$\ell\leq n$, pick a finite representative $T_\ell\in K_\ell$ (using
Proposition~\ref{prop:fmp}). It follows from
Lemma~\ref{lem:finite-width} that there is a $k\in\mathbb{N}$ such that,
for all $\ell\leq n$: ``$\phi$ can only count 
$T_\ell$-successors up
to $k$'', and hence, by $\sim^\phi_{i+1}$-equivalence, ``$\phi$ can
only count $K_\ell$-successors up to $k$, at depth $i$''. But then, it
follows that there are at most $k^n\cdot 2^{PROP(\phi)}$ many
$\sim^\phi_i$-equivalence classes. Moreover, all these classes are 
explicitly definable
by $GML$-formulas (in fact, by Boolean combinations of proposition
letters and formulas of the form $\Diamond_m\psi$ with $m\leq k$ and
$\psi$ a $GML$-formula defining some $\sim^\phi_{i+1}$-equivalence
class). 

   Thus, $\cL$ is not more expressive than $GML$ on tree models. It follows that
   $\cL$ is not more expressive than $GML$ on arbitrary Kripke models:
   consider any $\cL$-formula $\phi$, and let $\psi$ be any $GML$-formula
   equivalent to $\phi$ on tree models. If $\phi\leftrightarrow\psi$ were
   falsifiable on an arbitrary model, then, by unraveling, it could be falsified on a tree,
   which, by assumption, is not the case. Thus, $\phi$ and
   $\psi$ are equivalent on all Kripke models.
\end{proof}

\subsection{Graded modal logic on trees}

In this section, rather than assuming invariance for tree unraveling, we
consider only tree models from the start. That is, we view $GML$ as a
language for describing nodes of tree models. From this perspective,
$GML$ has three distinctive limiting features: \emph{(i)} when evaluated
in a node, formulas can only see the subtree starting from that node;
\emph{(ii)} when evaluated at a node, each formula can only look
finitely deeply into the subtree starting at that node; \emph{(iii)}
each formula can only count successors up to a finite natural number
(depending on the largest index occurring in the formula). 

\begin{fact}
On trees, $GML$ has the Compactness and L\"owenheim-Skolem properties.
\end{fact}

\begin{proof}
   This follows from the fact that $GML$ has these properties on arbitrary
   structures (recall that the tree unraveling of a countable model
   is still countable).
\end{proof}

We will turn this into a Lindstr\"om characterization for $GML$ on
trees. However, before we proceed, two technical issues need to be
discussed.

The first concerns \emph{closure under
   relativisation}. Recall from Sect.~\ref{sec:Lindstrom-ML} that all
abstract modal logics are assumed to satisfy this property. But what
does it mean for a logic to be closed under relativisation if we
consider only trees? Note that a submodel of a tree is not necessarily
a tree. We solve this problem as follows. Given a tree model $T$
containing a node $n$, and a unary predicate $p$ true at $n$, we
define $Subtree(T,n,p)$ to be the largest subtree of $T$ that contains
$n$ and contains only nodes satisfying $p$. Note that $n$ is not
necessarily the root of $Subtree(T,n,p)$. We say that an abstract
modal logic $\cL$ is closed under relativisation on trees, if for
every formula $\phi\in \cL$ and proposition letter $p$, there is a
formula $\psi\in \cL$ such that for all pointed tree models $(T,n)$,
$(T,n)\models \psi$ iff $(T,n)\models p$ and
$(Subtree(T,n,p),n)\models\phi$. In the case of $GML$, we can simply
pick $\psi$ to be the syntactic relativisation of $\phi$ by $p$, i.e.,
the formula obtained from $\phi\land p$ by replacing all subformulas
of the form $\Diamond_n\psi$ by $\Diamond_n(p\land\psi)$.

Secondly, we need to make an extra assumption, namely that the
extensions $\cL$ we consider are \emph{closed under substitution}.
Intuitively, this means that $\cL$ allows us to uniformly substitute
formulas for proposition letters. More precisely, $\cL$ is closed
under substitution if for all formulas $\phi, \psi\in \cL$ and
proposition letters $p$, there is a formula $\chi$ such that for all
pointed (tree) models $(M,w)$, $(M,w)\models\chi$ iff $(M^{[p\mapsto
  \{v\mid (M,v)\models\psi\}]},w)\models\phi$, where $M^{[p\mapsto
  \{v\mid (M,v)\models\psi\}]}$ is obtained from $M$ by changing the
valuation of $p$ to $\{v\mid (M,v)\models\psi\}$.  If we would not
assume closure under substitution, there would be proper extensions of
$GML$ with Compactness and L\"owenheim-Skolem. Indeed, the extension
of $GML$ with formulas of the form $\Diamond^{-}p$ (``the current node
has a parent that satisfies $p$'') for $p$ an atomic proposition, and
closed under the Boolean connectives, is an example. This logic is not
closed under substitution, as $p$ may not be replaced by a complex
formula in $\Diamond^{-}p$.

\begin{thm}
   \label{thm:Lindstrom-$GML$-trees}
   Let $\cL$ be an abstract modal logic closed under substitution and
   extending $GML$ on trees. $\cL$ satisfies
   Compactness and the L\"owenheim-Skolem property on trees iff
   it is no more expressive than $GML$ on trees.
\end{thm}

This is remarkable, since Compactness and L\"owenheim-Skolem are
also the characterizing features of first-order logic in the
classic Lindstr\"om theorem. Note that
first-order logic lacks Compactness on trees (due to the
connectedness of trees).

Now for the proof.  Let $\cL$ be an abstract modal logic satisfying
the conditions of Theorem~\ref{thm:Lindstrom-$GML$-trees}.
Lemma~\ref{lem:gen-subm-trees} below shows that $\cL$-formulas can only look
downwards in the tree.

\begin{lem} \label{lem:$GML$-tree-lemma} If $\cL$ is not
   invariant for generated submodels, then there is a formula
   $\chi\in \cL$ containing a unary predicate $p$ such that the following
   two conditions hold:

\begin{enumerate}[\em(1)]%\setlength{\itemsep}{0mm}
\item for all pointed trees $(T,n)$, $(T,n)\models\chi$ implies that
   $n$ has a parent satisfying $p$
\item there is a pointed tree $(T,n)$ satisfying $\chi$ in which $p$ is
    true only at the parent of $n$.
\end{enumerate}
\end{lem}

\begin{proof}
   Since $\cL$ is not invariant for generated submodels, there is a
   formula $\phi\in \cL$ and a pointed tree $(T,n)$ such that
   $(T,n)\models\phi$ and $(T_n,n)\not\models\phi$, or vice versa.
   Since $\cL$ is closed under negation, we may assume w.l.o.g.~that
   the former applies. Moreover,
   since $\cL$ is closed under renamings and the set of proposition
   letters is infinite, we may assume that there are
   infinitely many proposition letters not occurring in $\phi$, in the
   sense that their interpretation does not influence truth of $\phi$ at
   any state. We will refer to these proposition letters as
   being ``fresh''.

Pick two distinct fresh proposition letters $p,q$, let
   $\phi^p$ and $\phi^q$ be the relativisations of $\phi$ to $p$ and to
   $q$, and let $\Sigma$ be the following set of $\cL$-formulas:
   \[ \Sigma = \{\phi^p, \neg\phi^q, \Box^k (p\land q)\mid k\geq 0\}\]
   Observe that $\Sigma$ is satisfiable: it is true at $(T,n)$
   when we make $p$ true at all nodes, and $q$ only at $n$ and its
   descendants. We claim that truth of $\Sigma$ at a node $n$ in a
   tree implies that $n$ has a parent and it satisfies either $p$ or
   $q$. For, if not then the submodels $Subtree(T,n,p)$ and
   $Subtree(T,n,q)$ would coincide, and hence $\phi^p$ and $\phi^q$
   would have to have the same truth value at $n$.

   Next, we will use Compactness to obtain a finite subset of $\Sigma$
   that implies that the current node has a parent satisfying $p\lor
   q$. First, we `redescribe' the situation encoded by $\Sigma$ from
   the perspective of the parent node. Let $\Sigma'$ be the following
   set of $\cL$-formulas, where $r$ is another fresh proposition letter:
   \[ \Sigma' = \{\Diamond(\phi^p\land\neg\phi^q\land r), 
\Box(r\to\Box^k(p\land q))\mid k\geq 0\}\]
   By the previous observations, $\Sigma'\models p\lor q$.  Hence, by
   Compactness, there is a $\ell\in \mathbb{N}$ such that
   $\Diamond(\phi^p\land\neg\phi^q\land r) \land \bigwedge_{k\leq \ell}
   \Box(r\to\Box^k(p\land q))\models p\lor q$. Going back to the
   perspective of the node $n$, if we define $\psi$ to be the formula
   $\phi^p\land\neg\phi^q\land\bigwedge_{k\leq \ell}\Box^k(p\land q)$, then
   $\psi$ is satisfiable, and it implies the existence of a parent node
   satisfying $p\lor q$. 
   To see this, suppose for the sake of contradiction that $\psi$ is
   satisfiable in a pointed tree $(T',n')$ and $n'$ does \emph{not}
   have a parent satisfying $p\lor q$. There are two cases. If $n'$
   has a parent (satisfying $\neg p \land \neg q$), we immediately
   derive a contradiction, by our earlier arguments. If $n'$ does not
   have a parent, i.e., is the root of $T'$, we may change the tree by
   adding a parent above $n'$ satisfying $\neg p\land\neg q$, without
   affecting the truth of $\psi$ at $n'$, as follows from the
   definition of $\psi$ and the way we defined closure under
   relativisation. Therefore, we obtain again a contradiction.

   Finally, we take two more fresh proposition letters, $s$ and $t$,
   and we use the fact that $\cL$ is closed under substitution: we define
   $\chi$ to be $s\land \psi[p/(p\land (\Diamond s\to
   t)),q/(q\land(\Diamond s\to t))]$.

   On the one hand, truth of $\chi$ at a node implies it
   has a parent satisfying either $(p ~ \land ~ (\Diamond s\to t))$ or
   $(q ~ \land ~ (\Diamond s\to t))$, and hence $t$. On the other hand,
   there is a pointed tree satisfying $\chi$ in which $t$ is only
   true at the parent node: take $(T,n)$ and
   extend the valuation by making $s$ true only at $n$, and $t$
   at its parent.
\end{proof}

\begin{lem}\label{lem:gen-subm-trees}
   $\cL$ is invariant for generated submodels.
\end{lem}

\begin{proof}
   Suppose not. Let $\chi(p)\in\cL$ be as described by
   Lemma~\ref{lem:$GML$-tree-lemma}. By a ``fresh renaming'' of $\chi$ we
   will mean a copy in which all proposition letters have been
   renamed to fresh ones, and which has been relativised by an
   additional fresh proposition letter. For the reasons explained in
   the proof of Lemma~\ref{lem:$GML$-tree-lemma}, we may assume that
   $\chi(p)$ has infinitely many fresh renamings
   $(\chi_i(p_i))_{i\in\mathbb{N}}$.

   Finally, we define $\Sigma$ to be the set of $\cL$-formulas
   $\{\chi_1(p_1), \chi_1(\chi_2(p_2)), \chi_1(\chi_2(\chi_3(p_3))),
   \ldots\}$.  Every finite subset of $\Sigma$ is satisfiable.  Indeed,
   a satisfying model may be constructed by ``overlaying'' different
   copies of the model $(T,n)$ from Lemma~\ref{lem:$GML$-tree-lemma}, clause 2.
   On the other hand, if a node would satisfy all formulas in $\Sigma$
   at once, its ancestors would form an infinite ascending chain, which
   contradicts the well-foundedness property of trees.
\end{proof}

The remainder of the proof of Theorem~\ref{thm:Lindstrom-$GML$-trees} is
along the same lines as in Sect.~\ref{sec:$GML$-arbitrary}: first we
prove that $\cL$ has the finite occurrence, finite depth and finite
width properties on trees (using the fact that it is invariant for
generated submodels), and then we derive the Lindstr\"om theorem by
the same argument used in the proof of Theorem~\ref{thm:lindstrom-$GML$}.

\subsection{The guarded fragment}

The \emph{guarded fragment} $GF$ forms a second extension of modal
logic, incomparable to graded modal logic. It allows for arbitrary
quantifications of the form $\exists \vec{y}(G(\vec{x},\vec{y}) \land
\phi(\vec{x},\vec{y}))$, where $\vec{x}$ and $\vec{y}$ are tuples of
variables, and the \emph{guard} $G$ is an atomic formula containing
the variables in $\vec{x}$ and $\vec{y}$.  The guarded fragment is
decidable and has many `modal' meta-properties,
due to its invariance for \emph{guarded bisimulations}
\cite{Andreka98:modal,benthem2005:guards}, see
below for the definition.

Because of its modal character, $GF$ seems an obvious case for a
Lindstr\"om-style analysis like the one we have given for modal logic
and graded modal logic. However, there are some technical
difficulties, and we have not been able to prove an analogue of
Theorem~\ref{thm:Lindstrom-ML} for the guarded fragment yet. In this
section, we focus on a special case, in fact the same special case as
in the first half of the paper, namely for vocabularies with only
unary and binary predicates. The \emph{Binary Guarded Fragment}
($GF_\bin$) has the following syntax:
\[ \phi ::= R\vec{x} \mid x=y \mid \neg\phi \mid \phi\lor\psi \mid
         \exists y(G(x,y)\land\psi(x,y))
\]
where $R$ is a unary or binary atomic predicate, $x$ and $y$ are
distinct variables, the \emph{guard} $G$ is an atomic formula
containing both $x$ and $y$ (in any order) and $\psi$ contains no free
variables besides (possibly) $x$ and $y$.  Note that, by this
definition, unary guards such as $Py$, $Ryy$ and $y=y$ are not
allowed, and also unguarded quantification over a single remaining
free variable, as in $\exists y.\phi(y)$, can no longer be expressed
by $\exists y.(y=y \land \phi(y))$. Thus, every $GF_\bin$-formula
contains at least one free variable. This prohibition on unary guards
will be important, as it implies that the truth of a formula only
depends on a local neighborhood of the elements assigned to the free
variables.  

$GF_\bin$ can be seen as an abstract logic contained in $FO^2$,
provided that the definition of abstract logics is adapted in order to
take into account formulas with free variables, in the natural way.
Below, we will assume such a generalized notion of abstract
logic. When interpreted over Kripke models (and considering only
formulas with one free variable), the language $GF_\bin$ also constitutes an
abstract modal logic extending basic modal logic. 

Guarded bisimulations admit a natural adaptation to
this restricted version of the guarded fragment, which we call
$GF_\bin$-bisimulations, defined below.

\begin{thm}\label{thm:Lindstrom-GFbin}
   An abstract logic extending $GF_\bin$  satisfies Compactness
   and Invariance for $GF_\bin$-bisimulation iff it is no more
   expressive than $GF_\bin$.
\end{thm}

The proof is along the same lines as for modal logic and graded modal
logic: using Compactness, we prove a finite occurrence property and a
finite depth property (where depth is now measured as distance in the
`Gaifman graph'). We then use a tree-unraveling argument to show that
$GF_\bin$ can express all properties invariant for guarded
bisimulations and having the finite depth property. It is exactly in
this last step that the restriction to unary and binary relation
symbols, as well as the restrictions on the allowed guards, turn out
to be crucial. Roughly, these restrictions allows us to relate
distance in the unraveled tree to Gaifman distance in the original
structure. 

\medskip

In the remainder of this section, we present the proof in more detail.
We start with some modal features that hold for $GF$ in general.
First, there is a natural syntactic notion of formula depth, whose
inductive definition counts the above polyadic quantifiers as single
steps:
\[\begin{array}{@{}l@{~}l@{~}l@{}}
   depth(Px) &=& 0 \\
   depth (\neg\phi) &=& depth(\phi) \\
   depth(\phi\lor\psi) &=& \max (depth (\phi), depth (\psi)) \\
   depth(\exists\vec{y}(G(\vec{x},\vec{y}\land\phi(\vec{x},\vec{y}))) 
&=& depth(\phi)+1
   \end{array}\]

   Call a set of elements of $M$ ``\emph{guarded}'' if some tuple
   belonging to some atomic relation contains all these elements. We
   define distance for points in models $(M, \vec{s})$, where
   $\vec{s}$ is a tuple of nodes, as follows: $dist(\vec{s}, s_i, 0)$
   holds for $s_i \in \vec{s}$, and $dist(\vec{s}, t, n+1)$ holds if
   there is a $u$ with $dist(\vec{s}, u, n)$ and $\{t,u\}$ is
   guarded. Note that, by this definition, $dist$
   describes ``distance at most'' rather than ``exact distance''. 

We write $Cut((M, \vec{s}), n)$ for the submodel $\{t \in (M, \vec{s})
\mid dist(\vec{s}, t, n)\}$ consisting of all points $t$ in $M$ lying at
distance at most $n$ from $s$. The following result shows that $GF$,
like  basic modal logic, satisfies a \emph{finite depth
   property}, suitably defined:

\begin{lem}[Distance-Depth Lemma]
   Let $\phi$ be any guarded formula of depth $n$, and let $(N,
   \vec{s})$ be any submodel of $(M, \vec{s})$ containing all of
   $Cut((M, \vec{s}), n)$.  Then $(M, \vec{s}) \models \phi$ iff
   $(N, \vec{s}) \models \phi$.
\end{lem}

Next comes a generalization of modal bisimulation. A guarded
bisimulation is a non-empty set $F$ of finite partial isomorphisms
between two models $M$ and $N$ which has the following back-and-forth
conditions: given any function $f:X\to Y$ in $F$, (i) for all guarded
$Z\subseteq M$, there is a $g\in F$ with domain $Z$ such that $g$ and
$f$ agree on the intersection $X\cap Z$, (ii) for all guarded
$W\subseteq N$, there is a $g\in F$ with range $W$ such that the
inverses $g^{-1}$ and $f^{-1}$ agree on $Y\cap W$.

Also, we have ``\emph{rooted}'' guarded bisimulations $F$ running
between models $(M, s_1, \ldots, s_n)$ and $(N, t_1, \ldots, t_n)$
with given initial objects, where one requires that $\{(s_i,t_i)\mid
1\leq i\leq n\}$ is a partial isomorphism in $F$. By a simple
inductive argument, one then shows:

\begin{fact}
   $GF$-formulas are invariant for rooted guarded bisimulations:
   whenever there is a rooted guarded bisimulation between models
   $(M, s_1,\ldots,s_n)$ and $(N, t_1,\ldots,t_n)$, then
   for all $GF$-formulas $\phi(x_1,\ldots,x_n)$,
   $M\models\phi~[s_1,\ldots,s_n]$ iff $N\models\phi~[t_1,\ldots,t_n]$.
\end{fact}

Andr\'eka, van Benthem and N\'emeti \cite{Andreka98:modal} show that, 
conversely,
$GF$ consists, up to logical equivalence, of just those first-order
formulas which are invariant for guarded bisimulations.

Another `modal' use of guarded bisimulation in the same paper is \emph{model
unraveling}. This is like standard modal unraveling, but the
construction is a bit more delicate:

\begin{defi}[Tree unravelings for GF]
Let $M$ be any model. A \emph{guarded path} will be a non-empty finite
sequence of guarded sets in $M$. We say that an element of $M$ is
\emph{active} in a guarded path $\pi = \langle S_1, \ldots,
S_n\rangle$ if it belongs to $S_n$ and not to $S_{n-1}$ (or
$n=1$), and \emph{passive} if it belongs to both $S_n$ and $S_{n-1}$. 
The intuition is when an element occurs actively, a new copy is 
created in the unraveling, whereas when it occurs passively, the 
previous copy is used. Correspondingly, let $\equiv$ be the 
equivalence relation generated by $(\pi_1,d)\sim(\pi_2,d)$ for 
$\pi_2$ a guarded path extending $\pi_1$ with a single guarded set,
and $d$ passive in $\pi_2$. Now, the domain of the tree unraveling 
$\unr_{GF}(M)$ consists
of all pairs $(\pi,d)$ where $\pi$ is a guarded path and $d$ is active
in $\pi$, and the interpretation of predicate symbols $Q$ 
 is as follows. $I(Q)$ contains the tuple
 $\langle(\pi_1, d_1),\ldots,(\pi_k,d_k)\rangle$ if and only if
$(d_1, \ldots, d_n)$ belongs to the interpretation of $Q$ in $M$, and
there is a guarded path $\pi$ 
(which will in fact be a common extension of $\pi_1, \ldots, \pi_n$), 
such that $d_1, \ldots, d_k$ all belong to the final
set of $\pi$, and $(\pi,d_i)\sim (\pi_i,d_i)$ for all $1\leq i\leq n$. 

   For a model with parameters $(M, \vec{s})$ this generalizes as
   follows. $\unr_{GF}(M, \vec{s})$ has paths $\pi$ all starting from the
   initial set $\vec{s}$, but then continuing with guarded sets
   only. The objects $(\pi, d)$ are defined as before.
\end{defi} 
   The point here is that the set $F$ of all restrictions of the finite
   maps sending $(\pi_i,d_i)$ to $d_i$ for all guarded finite domains
   in $\unr_{GF}(M, \vec{s})$ is a rooted guarded bisimulation between $(M,
   \vec{s})$ and $\unr_{GF}(M, \vec{s})$. Checking the zigzag conditions
   for the bisimulation will reveal the reason for the above
   definition of the predicate interpretations $I(Q)$.
   For other, essentially equivalent formulations of this notion,
   cf.~\cite{Andreka98:modal,Graedel99:restraining}.

\medskip

   Now we have the generalities in place for our Lindstr\"om Theorem,
   but it remains to make some adjustments.
   Firstly, the notion of guarded bisimulation needs to be
   slightly adjusted. \emph{$GF_\bin$-bisimulations} are defined
   like guarded bisimulations, except that the back-and-forth
   conditions are only required to hold when $X\cap Z\neq\emptyset$ and
   $Y\cap W\neq\emptyset$, respectively. It can be shown that
   $GF_\bin$ is precisely the $GF_\bin$-bisimulation invariant fragment
   of $FO$ (over vocabularies consisting of unary and binary relation
   symbols only).

   Secondly, %
   in the
   definition of tree unravelings, we make one simple change for the
   binary case:
   \begin{quote}
   The finite paths of guarded sets always introduce one new object at
   each stage.  At each step, one takes a new object related
   to that new object.
   \end{quote}

   This allows paths starting with object $a$ and then continuing with
   $Rab$, $Qcb$, \ldots, while ruling out paths like $Rab, Qac$. But
   the final atom is not omitted from the unraveled model, since one
   can have paths starting with $a$ and then placing $Qac$
   immediately. Thus, even with these restricted paths, we still have a
   $GF_\bin$-bisimulation between tree unravelings and their original
   models. The real point of this adjustment is the following observation:

   The definition of predicates for path objects makes
   binary relations hold only between objects $(\pi_1, d_1), (\pi_2, d_2)$ where
   $\pi_2$ is a one-step continuation of the path $\pi_1$, or vice versa. But
   then, counting distance as before,
   \begin{quote}The new object at the end of a path of length $k$ lies
     at distance $k$ from the initial object of the path.
   \end{quote}

   Put in more vivid terms, `\emph{tree distance is true distance}' in
   the original model. This is a non-obvious fact. E.g., with ternary
   guards $Rayz$, objects at the end of a path may keep links to the
   initial object $a$ which might recur at any finite depth in the
   guarded sets along the path.\footnote{This observation is due to
     Martin Otto, p.c.~to the authors.}  Having a direct
   correspondence between the length of the path and the distance in
   the tree unraveling is essential, as it will allow us to establish
   a finite depth property analogous to Lemma~\ref{lem:ml-finite-depth}
   for the basic modal language. 

   We will write $\unr_{GF_\bin}(M)$ and
   $\unr_{GF_\bin}(M,\vec{s})$ for this new type of unraveling. Now 
everything is in place for our final argument:

\begin{proof}[Proof of Theorem~\ref{thm:Lindstrom-GFbin}]
   Let $\cL$ be any abstract logic extending $GF_\bin$ and satisfying
   Compactness and invariance for $GF_\bin$-bisimulations.
   As before, it suffices to show that every formula $\phi\in \cL$
   is invariant for models that are equivalent for all $GF_\bin$-formulas
   up to some finite depth $n$.

   First, largely as in the earlier modal proof of
   Sect.~\ref{sec:Lindstrom-ML}, we use the Compactness of $\cL$, together
   with its \emph{Relativization Closure}, to show that $\phi$ must
   have the \emph{Finite Occurrence Property} and a
   \emph{Finite Distance Property} for some level $n$. Before,
   universal prefix formulas $\Box^kp$ (for all finite $k$) made sure
   that $p$ holds in the generated submodel at the current node. This
   time, one uses all nested sequences of universal guarded quantifiers
   up to depth $k$, requiring that some new predicate $P$ holds for all
   objects reached at the end. The $n$ thus found for the local depth
   of the formula $\phi$ is the same $n$ as needed for the following
   semantic invariance:

   Given the above unraveling construction and invariance for
   $GF_\bin$-bisimulation for $\cL$, we may then assume, without loss of
   generality, that we have the following situation:

   \begin{enumerate}[(a)]
   \item $\unr_{GF_\bin}(M, \vec{s}) \models \phi$
   \item $\unr_{GF_\bin}(M, \vec{s})$ and $\unr_{GF_\bin}(N, 
\vec{t})$ satisfy the same
      $GF_\bin$-formulas up to depth $n$
   \end{enumerate}
Our aim is to show that $(\unr_{GF_\bin}(N, \vec{t})) \models\phi$.

We start by cutting the tree models to depth $n$, as  before in our
modal argument, obtaining $Cut(\unr_{GF_\bin}(M, \vec{s}), n)$ and 
$Cut(\unr_{GF_\bin}(N, \vec{t}),
n)$. Since tree depth is true depth, this does not change truth values
of $\phi$ in either model.

Then we define a $GF_\bin$-bisimulation between these models.
%$Cut(\unr_{GF_\bin}(M, \vec{s}), n)$ and $Cut(\unr_{GF_\bin}(N,
%\vec{t}), n)$. 
First of all, note that all elements in the first one
are of the form $(\langle \{\vec{s}\}, \{d_1, d_2\},
\ldots, \{d_{k-1}, d_k\}\rangle, d_k)$ with $d\in \vec{s}$ and $0\leq 
k\leq n$) where each set in
the sequence is guarded. Likewise for the second model.
Secondly, note that the only guarded subsets in these models are
singleton sets and sets of the form
$\{(\pi,d),(\langle\pi,\{d,e\}\rangle,e)\}$.
Now, let the relation $Z$ be the set consisting of 
$(\{\vec{s}\},\{\vec{t}\})$, as well
as all sets of pairs $$\{(\{(\pi,d),(\pi',d')),
((\langle\pi,\{d,e\}\rangle,e),(\langle\pi',\{d',e'\}\rangle,e'))\}$$
with $\pi, \pi'$ sequences of some length $k\leq n$ for which it 
holds that, for
all $i\leq k$, the guarded sets $\pi_i$ and $\pi'_i$ satisfy the same
$GF_\bin$-formulas of quantifier depth $n-i$.

It can be shown (the argument involves a case distinction)
that $Z$ is a $GF_\bin$-bisimulation between $Cut(\unr_{GF_\bin}(M,
\vec{s}), n)$ and $Cut(\unr_{GF_\bin}(N, \vec{t}), n)$.  Hence,
  the truth of $\phi$ transfers
from $Cut(\unr_{GF_\bin}(M, \vec{s}), n)$ to $Cut(\unr_{GF_\bin}(N,
\vec{t}), n)$, and hence to $(N, \vec{t})$.
\end{proof}

As indicated, this argument hits a barrier when ternary predicate are
allowed, and hence we leave a similar Lindstr\"om characterization of
the full guarded fragment $GF$ as an open problem. Recently
\cite{OttoPiro}, a Lindstr\"om-style characterization was obtained for
the full $GF$ along slightly different lines, by means of adding a
further model-theoretic property, cf.~the discussion below.

\section{Discussion}

To conclude, we identify a few lines of research suggested by our results.

\subsection{Charting the landscape of first-order fragments}
We have seen how many fragments of first-order logic have Lindstr\"om 
characterizations, but with very different proof techniques. 
`Top-down', we showed how careful coding can get the original proof 
down to cover the 3-variable fragment, while `bottom up', we gave a 
modal technique which lifts to various richer modal-like extensions. 

There are more positive results than we have presented here: e.g., our
coding technique can easily be adapted to the \emph{bounded fragment}
of first-order logic (cf.~\cite{cate04}), showing that \emph{every
  abstract logic extending the bounded fragment and having Compactness
  and the L\"owenheim-Skolem property is contained in FO}, and hence
also \emph{the bounded fragment is maximal with respect to
  Compactness, the L\"owenheim-Skolem property, and invariance for
  generated submodels} (a restriction to binary vocabularies is not
necessary here).

However, there is also a `gap' between the two proof techniques, and 
logics like the 2-variable fragment, or modal logic with an added 
`universal modality' (``true in all nodes, accessible or not'') pose a 
challenge. We may need new proof techniques here, or indeed, a 
rethinking of what a Lindstr\"om theorem should be in such cases. 
Recently, Otto and Piro \cite{OttoPiro} have characterized modal 
logic extended with the universal modality as being a maximal system 
$\cL$ with respect to Compactness, \emph{invariance for global 
bisimulations} and also, the \emph{Tarski union property} (i.e., 
preservation of $\cL$-sentences under limits of chains of $\cL$-elementary 
embeddings), and similarly for the guarded fragment. 
Thus, additional preservation properties may be needed.

\subsection{Re-positioning first-order logic in this broader setting}
Lindstr\"om's theorem in its classical formulation characterizes
first-order logic only within the class of its extensions. A natural
question to ask in our wider setting is within which broader classes 
of languages we can characterize first-order logic. Our 
Theorem~\ref{thm:strong-Lindstrom-FO} gave a partial answer: it 
characterized first-order logic within
the class of extensions of $FO^3$. This at once suggests many new 
questions. For instance, are there extensions of $FO^2$ not contained 
in $FO$ that satisfy
  Compactness and the L\"owenheim-Skolem property? \footnote{ 
Theorem~\ref{thm:well-behaved-extension} gave an extension of 
\emph{modal logic} not contained in first-order logic which has these 
properties.}

\subsection{Down and then up again: adding fixed-point operators}
It is well-known that fragments of first-order logic can behave very 
differently from the full system when it comes to adding transitive 
closure, or even complete fixed-point extensions. In particular, the 
\emph{modal $\mu$-calculus} remains decidable, just like the basic 
modal logic, whereas the logic $LFP(FO)$ extending first-order logic 
with arbitrary fixed-point operators becomes not recursively 
enumerable, and indeed $\Pi_{1}^1$ complete. While fixed-point logics 
of both sorts are natural from a computational point of view, they 
have resisted Lindstr\"om-style analysis so far. Here is one 
question, out of many which suggest themselves following our earlier results:

\begin{qu}
   Can the modal $\mu$-calculus be characterized in terms of
   bisimulation invariance and the finite model property?
\footnote{This formulation naturally arises in our setting since so many of our 
modal arguments involved finite reachability of nodes from the 
origin, a typical non-first-order fixed-point notion.}
\end{qu} 
\subsection{Characterizing logics on specific classes of structures.}
No Lindstr\"om characterizations are known for first-order logic on
finite structures, or on trees. Compactness fails for first-order
logic on such structures, and, on finite structures, the
L\"owenheim-Skolem property becomes meaningless. In this paper, we
proved one positive result: we showed that $GML$ behaves on trees as
first-order logic does on arbitrary structures: it is maximal with
respect to Compactness and the L\"owenheim-Skolem property.  In
general, however, this area remains underexplored:
\cite{Kolaitis95:generalized} raise the issue of the `missing 
Lindstr\"om theorem' for finite model theory. Here we state just one 
related technical question. Given a logic $\mathcal{L}$ interpreted
on finite structures and a logic $\mathcal{L'}$ interpreted on
arbitrary structures (both satisfying the usual conditions such as 
closure under the Boolean operations), 
let us say that $\mathcal{L'}$ is a conservative
extension of $\mathcal{L}$ if $\mathcal{L}$ is simply the restriction
of $\mathcal{L'}$ to finite structures.

\begin{qu}
   Are there extensions of first-order logic on finite structures
   that have a conservative extension to infinite structures
   satisfying Compactness and L\"owenheim-Skolem?
\end{qu}

\subsection{Relating different characterizations: the case of interpolation}

Finally, taking stock of our results, there is also the more elusive, 
but still interesting question of ``what is a Lindstr\"om theorem?''. 
As we already indicated in the introduction, there are many natural 
ways to
characterize the expressive power of a logic, including Lindstr\"om
theorems, preservation theorems, and characterizations involving
Craig interpolation.

And even the Lindstr\"om theorems in this paper fall into different 
classes. Some use Compactness and
L\"owenheim-Skolem, or some other size-restricting principle of model
existence, while others use Compactness and Invariance for Potential
Isomorphisms, or some other semantic transfer property of the
language. While these formulations both capture first-order logic, we
have seen that they diverge for basic modal logic, where we have a
characterization of the latter type but not of the former. The precise
extent of this phenomenon remains to be understood.

\medskip

\def\fol{\mbox{FO}}\def\sol{\mbox{SO}}
\def\phi{\varphi}\def\mm{\mathcal{M}}\def\I{\mbox{I}}\def\II{\mbox{II}}
\def\o{\omega}\def\k{\kappa}
\def\lhyp{L_{\mbox{\tiny HYP}}}
\def\loyo{L_{\o_1\o}}
\def\lio{L_{\infty\o}}

While the picture of `natural properties' is not quite clear yet,
it is noticable that Lindstr\"om theorems seem to go hand in hand
with \emph{interpolation results} in abstract model theory and related 
areas.  Among the proper extensions of $\fol$ the main examples of 
interpolation are $\loyo$, its countable
admissible fragments $L_A$, and second order logic $\sol$. Barwise
\cite{MR0337483} gives a maximality characterization of $\loyo$ and
its countable admissible fragments $L_A$ in terms of a (generalized)
recursion theoretic criterion called {\em strict absoluteness}. The
proof uses interpolation.\footnote{It is arguable whether this 
characterization
of $\loyo$ (or $L_A$) should be considered a Lindstr\"om type results
as strict absoluteness is not a model theoretic condition. However the
boundary between generalized recursion theory and infinitary model
theory is not very strict.}  Also second-order logic has a maximality
characterization: $\sol$ is the maximal extension of $\fol$ in which
every definable model class has a {\em ``flat''} definition in set
theory \cite{MR567682}. Again this result is intimately
connected with the (trivial) way in which $\sol$ satisfies
(single-sorted) interpolation.  Finally, the interpolation theorem of
$\lio$ for entailment along potential isomorphism in \cite{MR1777793}
is related to a maximality characterization of this logic: $\lio$ is
the maximal extension of $\fol$ which is bounded and has the Karp
Property \cite{MR0376337}. We feel that this link between Lindstr\"om 
theorems and interpolation theorems requires further analysis.

Moreover, this connection makes sense for our study of fragments as 
well. Some of our results used interpolation properties already, and 
in particular, our paradigmatic modal logics have them. Indeed, among 
fragments, even further types of interpolation property emerge, which 
fail for first-order logic. Thus, the two major modal logics for 
which we managed to obtain Lindstr\"om
theorems, namely basic modal logic and graded modal logic, both
possess \emph{uniform interpolation}, a strong form of Craig
interpolation where the interpolant can be chosen uniformly for all 
consequents sharing some specified vocabulary with the antecedent. We 
would like to understand the impact of this condition on abstract 
logics in conjunction with our Lindstr\"om-style analysis in this 
paper.

\bibliographystyle{plain}
\bibliography{balders-collection}

\end{document}